\documentclass[%
superscriptaddress,
nofootinbib,
natbib,
amsmath,
amssymb,
aps,
pra,
]{revtex4-2}

\usepackage[english]{babel}

\usepackage[letterpaper,top=2cm,bottom=2cm,left=3cm,right=3cm,marginparwidth=1.75cm]{geometry}

\usepackage{graphicx}
\usepackage{dcolumn}
\usepackage{bm}
\usepackage{bbold}
\usepackage{xcolor}

\usepackage{amsmath}
\usepackage{amsthm}
\usepackage{multirow}
\usepackage{colortbl}
\usepackage[normalem]{ulem}
\usepackage{etoolbox}

\usepackage{amssymb}

\usepackage[colorlinks=true, allcolors=blue]{hyperref}
\usepackage{braket}
\usepackage{tikz}
\usetikzlibrary{quantikz}

\begin{document}

\preprint{APS/123-QED}


\title{Efficient Multi-Controlled Gate Implementation in Trapped-Ion Systems}

\author{Minhyeok Kang}
\thanks{These authors contributed equally to this work.}
\affiliation{SKKU Advanced Institute of Nanotechnology (SAINT), Sungkyunkwan University, Suwon 16419, South Korea}

\author{Taejin Kim}
\thanks{These authors contributed equally to this work.}
\affiliation{Department of Chemistry, Yonsei University, Seoul 03722, Republic of Korea}

\author{Jungsoo Hong}
\affiliation{SKKU Advanced Institute of Nanotechnology (SAINT), Sungkyunkwan University, Suwon 16419, South Korea}

\author{Joonsuk Huh}
\email{joonsukhuh@yonsei.ac.kr}
\affiliation{Department of Chemistry, Yonsei University, Seoul 03722, Republic of Korea}
\affiliation{Department of Quantum Information, Yonsei University, Incheon 21983, Republic of Korea}

\begin{abstract}
Multi-controlled gates are essential primitives in quantum algorithms, yet implementing them via standard gate-level decompositions remains resource-intensive.
We develop efficient pulse-level implementations of multi-controlled gates in trapped-ion systems using the Cirac--Zoller scheme. 
We first show that the Cirac--Zoller construction admits a freedom in the sign choice of red-sideband (RSB) pulses, which leaves the logical operation invariant up to a local Pauli-\(Z\) correction. 
By exploiting this freedom, we construct equivalent realizations of multi-controlled gates and develop pulse cancellation for more efficient implementations of successive gates.
We perform numerical simulations and show that pulse cancellation reduces the gate time and improves the state fidelity.
Furthermore, we propose ancilla-free circuits for general \(N\)-controlled gates that use a single-controlled gate primitive and \(\mathcal{O}(N)\) RSB pulses.
As a key application, we apply our pulse cancellation to the linear combination of unitaries (LCU) method for block encoding.
We show that the RSB-pulse cost of the \textit{select} operator over \(L\) unitaries can be reduced from \(\mathcal{O}(L\log L)\) to \(\mathcal{O}(L)\), which improves the efficiency and scalability of LCU-based quantum circuits.
\end{abstract}

\maketitle


\section{Introduction}
\label{sec:intro}

Multi-controlled gates appear as core subroutines in a wide range of quantum algorithms, including Grover's search~\cite{grover1996}, quantum phase estimation~\cite{kitaev1995}, and linear combination of unitaries (LCU)-based Hamiltonian simulation~\cite{childs2012hamiltonian}.
Therefore, their efficient implementation directly governs the overall resource cost of these algorithms.
A standard approach is to decompose them into single-qubit gates and CNOTs~\cite{barenco1995elementary}, but this decomposition typically requires a number of elementary gates that grows rapidly with the number of control qubits.

For example, the \(N\)-Toffoli gate---which generalizes CNOT to \(N-1\) control qubits and one target qubit---admits a standard decomposition requiring \(\mathcal{O}(N^2)\) two-qubit gates without ancillary qubits, or \(\mathcal{O}(N)\) two-qubit gates together with \(\mathcal{O}(N)\) ancillary qubits~\cite{barenco1995elementary,Nielsen_Chuang_2010}.
Such overhead increases circuit depth and execution time, thus amplifying the effects of noise and decoherence on current hardware~\cite{Preskill2018quantum}.
These considerations motivate the search for efficient, hardware-native implementations of multi-controlled gates that avoid the overhead of generic elementary-gate decomposition.

\begin{figure}
    \centering
    \includegraphics[width=0.5\linewidth]{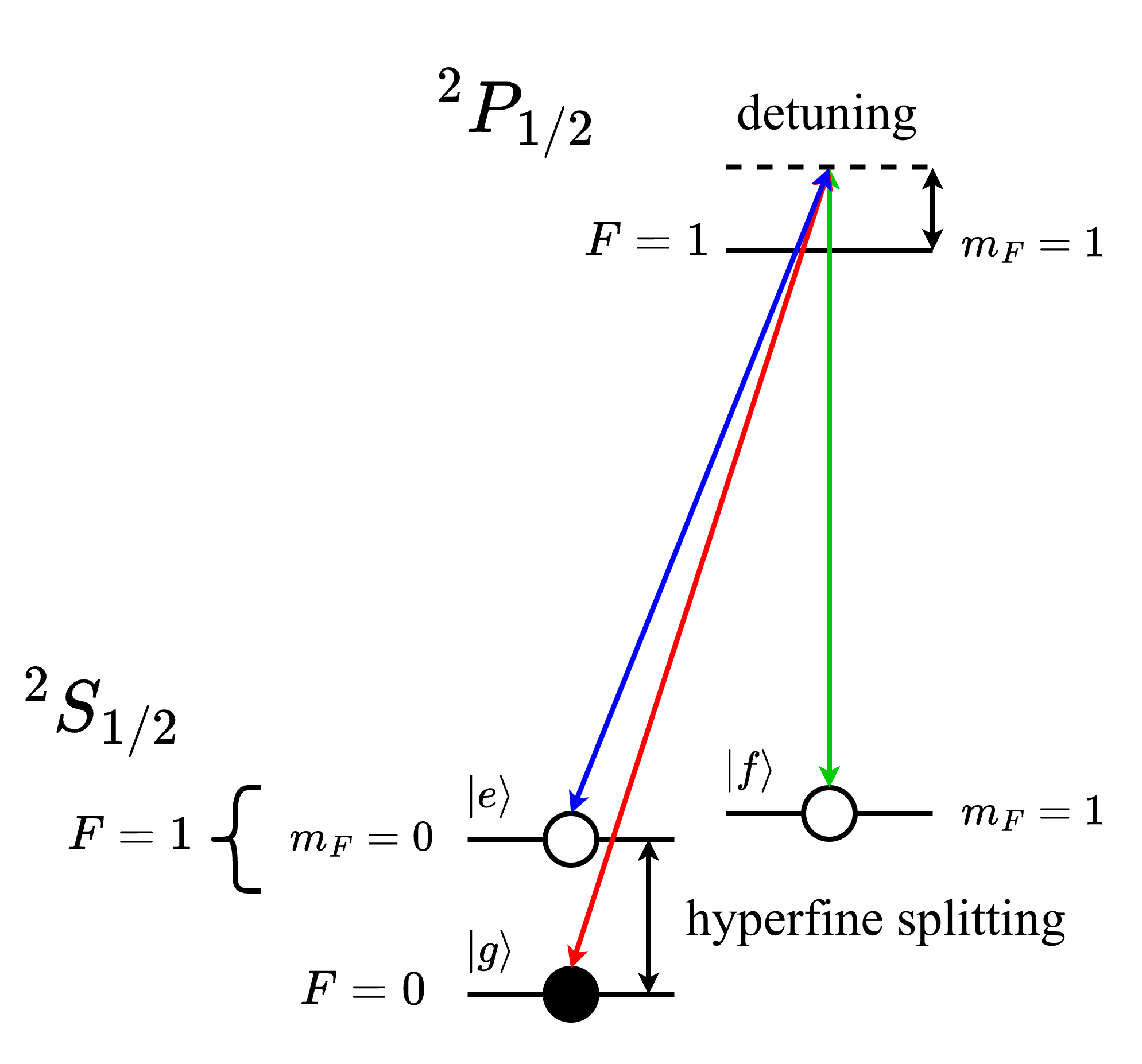}
    \caption{Energy level structure of the \(^{171}\mathrm{Yb}^+\) ion relevant to the Cirac--Zoller scheme.
    The hyperfine clock states define the qubit basis \(\ket{g}\) and \(\ket{e}\), while an additional Zeeman sublevel is used as the auxiliary state \(\ket{f}\).
    Two Raman beams couple the \({}^{2}S_{1/2}\) ground-state manifold via virtual excitation to the \(^{2}P_{1/2}\) level, inducing transitions \(\ket{g}\!\leftrightarrow\!\ket{e}\) and \(\ket{g}\!\leftrightarrow\!\ket{f}\).
    When tuned to the RSB, the Raman beams also drive spin--motion coupling with the vibrational mode.
    Adapted from Ref.~\cite{fang2023realization}.}
    \label{fig:Yb}
\end{figure}

Trapped-ion systems provide a natural setting for pulse-level control.
In these systems, each qubit is encoded in two computational basis states, denoted by \(\ket{g}\) and \(\ket{e}\), while collective vibrational modes of the ion chain serve as bosonic degrees of freedom that mediate entangling operations between ions.
In particular, the Cirac--Zoller scheme~\cite{cirac1995quantum} realizes multi-qubit gates through an auxiliary internal state \(\ket{f}\) and red-sideband (RSB) pulses~\cite{Leibfried2003quantum} that couple the internal and motional degrees of freedom.
As a concrete realization~\cite{fang2023realization}, in \({}^{171}\mathrm{Yb}^{+}\) the computational basis can be encoded in the hyperfine clock states \(\ket{g}=\ket{F=0,m_F=0}\) and \(\ket{e}=\ket{F=1,m_F=0}\) within the \({}^{2}S_{1/2}\) ground-state manifold, while the auxiliary state \(\ket{f}\) can be chosen from the Zeeman sublevels \(\ket{F=1,m_F=\pm1}\) that participate in the RSB coupling, as illustrated in Fig.~\ref{fig:Yb}.
Within this scheme, the \(N\)-Toffoli gate can be implemented using \(2N\) RSB \(\pi\) pulses and two single-qubit rotation gates, without ancillary qubits.
Recently, \citet{fang2023realization} experimentally demonstrated a \(5\)-Toffoli gate in a linear Paul trap with \({}^{171}\mathrm{Yb}^{+}\) ions.

Building on this Cirac--Zoller construction, \citet{kang2025doubling} proposed a circuit for a general \(N\)-controlled gate using \(2(N+1)\) RSB \(\pi\) pulses, a single-controlled gate, and one ancillary qubit initialized in \(\ket{g}\).
Given access to such a single-controlled gate, this construction achieves \(\mathcal{O}(N)\) pulse scaling and is more efficient than conventional gate decomposition. 
Since non-Clifford gates dominate the cost of fault-tolerant multi-controlled gate implementations~\cite{PhysRevA.71.022316,Campbell2017}, such pulse-level savings are also relevant for reducing resource overhead in the fault-tolerant regime.
Motivated by these pulse-level constructions, we study pulse-level implementations of multi-controlled gates in the Cirac--Zoller scheme, including the \(N\)-Toffoli gate, and explore how they can be further optimized.

In Sec.~\ref{sec:main_result}, we first show that the Cirac--Zoller implementation of the \(N\)-Toffoli gate admits multiple equivalent pulse-level realizations arising from different sign choices of the RSB pulses. 
We refer to this freedom in the sign choice of RSB pulses as \textit{gauge freedom}. 
By exploiting this gauge freedom, we implement the \(N\)-CSWAP (\(N\)-controlled SWAP) gate, which is decomposed into three consecutive \((N+2)\)-Toffoli gates~\cite{Nielsen_Chuang_2010}, with direct cancellation of adjacent inverse RSB subsequences.
Numerical simulations show that this pulse-cancelled implementation reduces the gate time and improves the fidelity under realistic noise. 
We then generalize this idea by proposing a new ancilla-free Cirac--Zoller construction for a general \(N\)-controlled gate that eliminates the ancillary qubit while retaining linear pulse scaling, using two single-qubit rotation gates instead.
The same RSB gauge freedom then yields efficient implementations of successive multi-controlled gates in this broader setting.

In Sec.~\ref{sec: LCU}, we consider the linear combination of unitaries (LCU) method~\cite{childs2012hamiltonian} as a key application. 
The LCU framework is widely used in quantum algorithms, including quantum signal processing (QSP)~\cite{PhysRevLett.118.010501,Kane2025blockencodingbosons} and quantum singular value transformation (QSVT)~\cite{10.1145/3313276.3316366}. 
Because the select operator in LCU is built from successive multi-controlled gates, our pulse cancellation applies directly to its implementation. 
We show that the RSB pulse cost of the select operator can be reduced from \(\mathcal{O}(L\log L)\) to \(\mathcal{O}(L)\), improving the efficiency and scalability of LCU-based circuits.

\section{Gauge freedom in Cirac--Zoller based multi-controlled gate construction}
\label{sec:main_result}

The Cirac--Zoller based scheme~\cite{cirac1995quantum,kang2025doubling} constructs entangling gates through RSB pulses that couple the internal states of the ions to a shared motional mode.
In this section, we show that there is a freedom in the sign choice of the RSB pulses, which leaves the logical operation invariant up to a phase that can be corrected by a single-qubit gate.
We refer to this freedom as \textit{gauge freedom}.
As a consequence, the number of required RSB pulses for successive multi-controlled gates can be reduced.

Recall that \(\ket{g}\) and \(\ket{e}\) denote the computational basis states of the internal qubit, and that \(\ket{f}\) denotes an auxiliary internal state.
The vibrational mode is represented by \(b\), with its Fock states labeled as \(\ket{n = 0,1,\dots}_b\).
We use two types of RSB transitions that couple the internal states to the shared vibrational mode: one between \(\ket{g}\) and \(\ket{e}\), and another between \(\ket{g}\) and \(\ket{f}\).
The corresponding RSB pulses, generated by the interaction Hamiltonians in Ref.~\cite{Leibfried2003quantum}, are
\begin{align}
    \label{eqn:rsb_unitary}
    \hat{U}_{\mathrm{RSB}}(\theta,\phi) = \exp \left(-\mathrm{i}\frac{\theta}{2}\left( \hat{a}\ket{e}\bra{g}\mathrm{e}^{\mathrm{i}\phi} + \hat{a}^{\dagger}\ket{g}\bra{e}\mathrm{e}^{-\mathrm{i}\phi} \right) \right),\\
    \label{eqn:rsb_aux_unitary}
    \hat{U}_{\mathrm{RSB}}^{\mathrm{aux}}(\theta,\phi) = \exp \left(-\mathrm{i}\frac{\theta}{2}\left( \hat{a}\ket{f}\bra{g}\mathrm{e}^{\mathrm{i}\phi} + \hat{a}^{\dagger}\ket{g}\bra{f}\mathrm{e}^{-\mathrm{i}\phi} \right) \right),
\end{align}
where \(\theta = t\Omega\eta\), with \(t\) the interaction time, \(\Omega\) the on-resonance carrier Rabi frequency, and \(\eta\) the Lamb-Dicke parameter.
The operators \(\hat{a}\), \(\hat{a}^{\dagger}\) are annihilation and creation operators of the vibrational mode, respectively.
For simplicity, we define \(\hat{U}_{\mathrm{RSB}}(\theta) = \hat{U}_{\mathrm{RSB}}(\theta,0)\) and \(\hat{U}_{\mathrm{RSB}}^{\mathrm{aux}}(\theta) = \hat{U}_{\mathrm{RSB}}^{\mathrm{aux}}(\theta,0)\).
Throughout this paper, we refer to both \(\hat{U}_{\mathrm{RSB}}(\theta)\) and \(\hat{U}_{\mathrm{RSB}}^{\mathrm{aux}}(\theta)\) collectively as RSB \(\theta\) pulses.

The RSB pulses we mainly use are \(\pi\) pulses and their inverses, \(-\pi\) pulses.
The action of \(\hat{U}_{\mathrm{RSB}}(\pm\pi)\) on the basis states that are relevant to our protocol is:
\begin{align}
    \label{eqn:rsb1}
    &\hat{U}_{\mathrm{RSB}}(\pm\pi)\ket{0}_b\ket{g} = \ket{0}_b\ket{g},\\
    \label{eqn:rsb2}
    &\hat{U}_{\mathrm{RSB}}(\pm\pi)\ket{0}_b\ket{e} = \mp \mathrm{i}\ket{1}_b\ket{g},\\
    \label{eqn:rsb3}
    &\hat{U}_{\mathrm{RSB}}(\pm\pi)\ket{1}_b\ket{g} = \mp \mathrm{i}\ket{0}_b\ket{e},
\end{align}
and the action of \(\hat{U}_{\mathrm{RSB}}^{\mathrm{aux}}(\pm\pi)\) on the basis states that are also relevant to our protocol is:
\begin{align}
    \label{eqn:rsbaux1}
    &\hat{U}_{\mathrm{RSB}}^{\mathrm{aux}}(\pm\pi)\ket{0}_b\ket{g} = \ket{0}_b\ket{g},\\
    \label{eqn:rsbaux2}
    &\hat{U}_{\mathrm{RSB}}^{\mathrm{aux}}(\pm\pi)\ket{0}_b\ket{e} = \ket{0}_b\ket{e},\\
    \label{eqn:rsbaux3}
    &\hat{U}_{\mathrm{RSB}}^{\mathrm{aux}}(\pm\pi)\ket{0}_b\ket{f} = \mp \mathrm{i}\ket{1}_b\ket{g},\\
    \label{eqn:rsbaux4}
    &\hat{U}_{\mathrm{RSB}}^{\mathrm{aux}}(\pm\pi)\ket{1}_b\ket{g} = \mp \mathrm{i}\ket{0}_b\ket{f},\\
    \label{eqn:rsbaux5}
    &\hat{U}_{\mathrm{RSB}}^{\mathrm{aux}}(\pm\pi)\ket{1}_b\ket{e} = \ket{1}_b\ket{e}.
\end{align}
The action of the RSB \(\pm\pi\) pulses described above corresponds to an energy exchange accompanied by a phase factor \(\mp \mathrm{i}\).

\subsection{Cirac--Zoller based scheme and gauge freedom in RSB pulses}

\begin{figure}[htb]
    \centering
    \includegraphics[width=\linewidth]{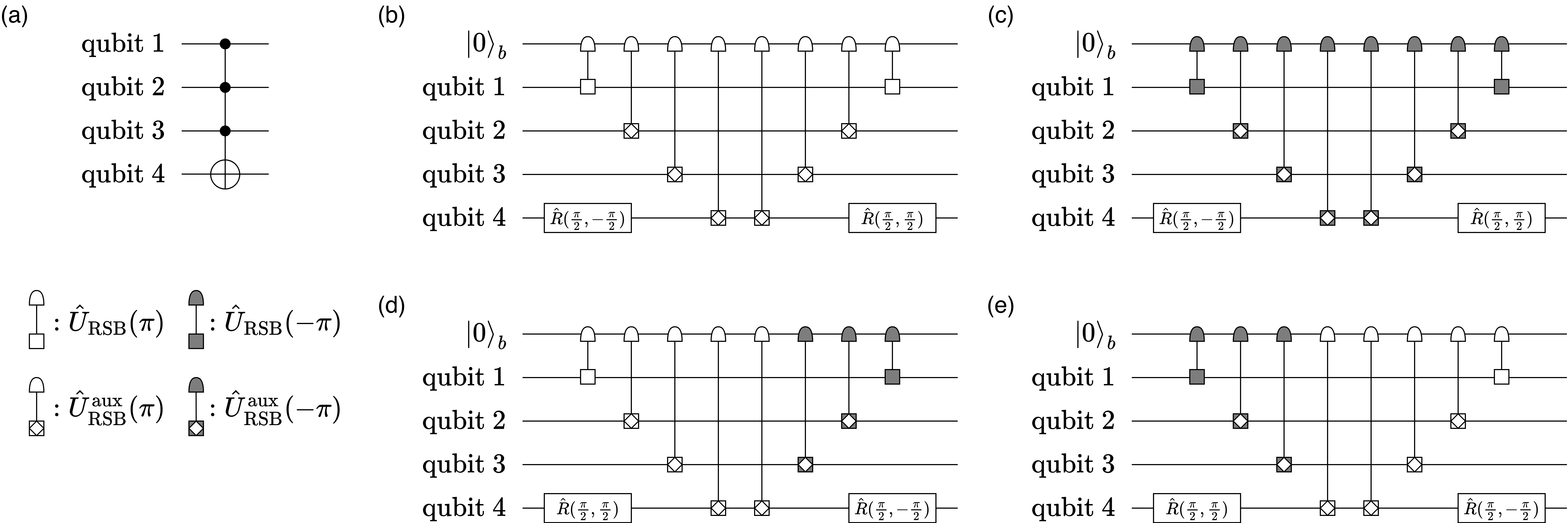}
    \caption{Implementation of the 4-Toffoli gate using the Cirac--Zoller scheme.
    (a) Circuit diagram of the 4-Toffoli gate.
    (b) Standard Cirac--Zoller scheme for 4-Toffoli gate, 
    constructed solely from RSB \(\pi\) pulses~\cite{fang2023realization}.
    (c)--(e) Alternative implementations obtained by exploiting the RSB gauge freedom, yielding equivalent operations with different pulse sequences.
    }
    \label{fig:toffoli}
\end{figure}

Consider the Cirac--Zoller scheme for the \(N\)-Toffoli gate.
The \(N\)-Toffoli gate (Fig.~\ref{fig:toffoli}~(a)) can be implemented using \(2N\) RSB \(\pi\) pulses along with two single-qubit rotation gates, as illustrated in Fig.~\ref{fig:toffoli}~(b).
A key idea is to encode the joint control condition into a phonon in the vibrational mode.
The first \(N\) RSB pulses conditionally excite a phonon depending on the state of the control qubits, and the two subsequent RSB pulses on the target qubit induce a relative phase that depends on the phonon occupation.
The last \(N\) RSB pulses then restore the control qubits to their initial states.
Two single-qubit rotation gates implement the phase-kickback mechanism; without them, the RSB pulse sequence realizes an \((N-1)\)-controlled \(Z\) gate.

To see this explicitly, consider the actions of two sequences of RSB pulses on the control qubits (the first and last \(N\) RSB \(\pi\) pulses).
When the control condition is not satisfied, the total number of energy exchanges is zero or four, and no phonon is present when the RSB pulses act on the target qubit.
Since each exchange contributes a phase factor of \((-\mathrm{i})\), zero or four exchanges yield the same accumulated phase, \((-\mathrm{i})^0 = (-\mathrm{i})^4 = 1\).
In contrast, when the control condition is satisfied, only two energy exchanges occur, giving a phase of \((-\mathrm{i})^2 = -1\), and a single phonon remains in the vibrational mode during the target-qubit operations.
This phonon induces an additional phase shift on the RSB pulses acting on the target qubit, which is precisely the mechanism by which the Cirac--Zoller scheme constructs the \((N-1)\)-controlled \(Z\) gate.

Notably, there is an RSB gauge freedom: we can apply not only RSB \(\pi\) pulses, but also \(-\pi\) pulses.
As presented in Fig.~\ref{fig:toffoli}~(c--e), the first or last (or both) sequence of RSB \(\pi\) pulses can be replaced with RSB \(-\pi\) pulses.
This gauge freedom arises because replacing RSB \(\pi\) pulses with \(-\pi\) pulses does not alter the energy exchange mechanism.
The only difference is the accumulated phase when the control condition is satisfied: Fig.~\ref{fig:toffoli}~(b--c) give \((-\mathrm{i})^2 = \mathrm{i}^2 = -1\), whereas Fig.~\ref{fig:toffoli}~(d--e) give \((-\mathrm{i})\cdot \mathrm{i} = 1\).
These phase differences can be corrected by appropriate single-qubit rotation gates: Fig.~\ref{fig:toffoli}~(b--c) uses \(\hat{R}(\pi/2,-\pi/2)\) first and \(\hat{R}(\pi/2,\pi/2)\) second, while Fig.~\ref{fig:toffoli}~(d--e) uses \(\hat{R}(\pi/2,\pi/2)\) first and \(\hat{R}(\pi/2,-\pi/2)\) second.

\begin{figure}
    \centering
    \includegraphics[width=1.0\linewidth]{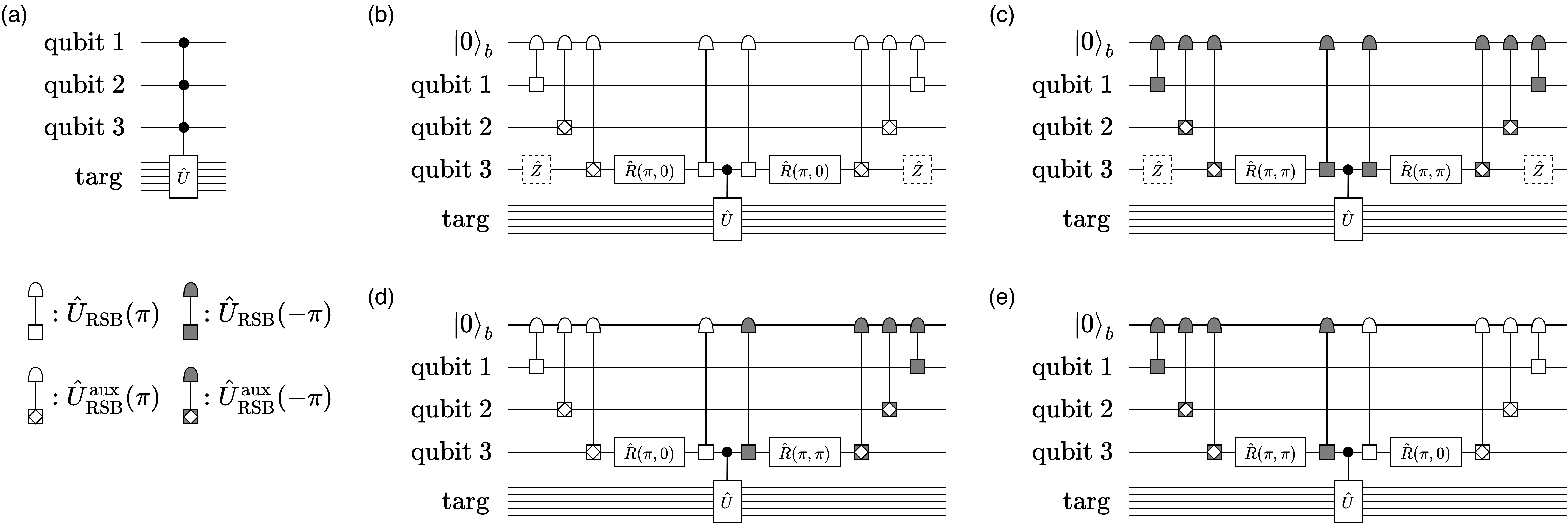}
    \caption{Implementation of a 3-controlled gate using the Cirac--Zoller based scheme.
    (a) Circuit diagram of a 3-controlled gate.
    (b--e) Decompositions of the circuit in (a) using four different types of RSB sequence variants.
    In (b) and (c), a \(Z\) gate (indicated by the black dashed box) is placed at one of two symmetric locations to compensate for the residual phase.
    In (d) and (e), where the encoding and decoding RSB sequences have opposite signs, the \(Z\) gate is not required.}
    \label{fig:mc_gate}
\end{figure}

\citet{kang2025doubling} proposed a Cirac--Zoller based scheme that extends single-controlled gates to multi-controlled gates with \(\mathcal{O}(N)\) RSB pulses.
This scheme introduces an ancillary qubit initialized in \(\ket{g}\), which absorbs the phonon that encodes the joint control condition and serves as the effective control for the single-controlled gate.
We propose an ancilla-free version of this scheme by redirecting the RSB pulses originally applied to the ancillary qubit onto the last control qubit and inserting \(\hat{R}(\theta,\phi)\) gates on both sides (Fig.~\ref{fig:mc_gate}).
Note that \(\hat{R}(\theta,\phi)\) acts only on the \(\{\ket{g},\ket{e}\}\) subspace and has no effect on the \(\ket{f}\) state:
\begin{gather}
    \hat{R}(\theta,\phi)
    =
    \begin{pmatrix}
    \cos\frac{\theta}{2} & -\mathrm{i} \mathrm{e}^{-\mathrm{i}\phi}\sin\frac{\theta}{2} \\
    -\mathrm{i} \mathrm{e}^{\mathrm{i}\phi}\sin\frac{\theta}{2} & \cos\frac{\theta}{2}
    \end{pmatrix}_{\{\ket{g},\ket{e}\}},\\
    \hat{R}(\theta,\phi)\ket{f} = \ket{f}.
\end{gather}
In addition, a \(Z\) gate is applied before or after the RSB sequence to match the resulting phase, up to a global phase.

The RSB \(\pi\) pulses in this circuit can also be replaced with the RSB \(-\pi\) pulses (Fig.~\ref{fig:mc_gate}~(c--e)).
In this case, the single-qubit rotation gate adjacent to the RSB pulse sequence is determined as follows: if the adjacent RSB pulse sequence is a \(\pi\) pulse sequence, the corresponding single-qubit rotation gate is \(\hat{R}(\pi, 0)\), whereas if it is a \(-\pi\) pulse sequence, the gate is \(\hat{R}(\pi, \pi)\).
Note that when the RSB pulse sequences on both sides are different, the \(Z\) gate can be omitted (Fig.~\ref{fig:mc_gate}~(d--e)).
The detailed derivation is given in Appendix~\ref{appendix:detailed_v2}.

\subsection{Pulse cancellation in successive multi-controlled gates}
\label{subsec:pulse_cancellation}

\begin{figure}[htb!]
    \centering
    \includegraphics[width=1.0\linewidth]{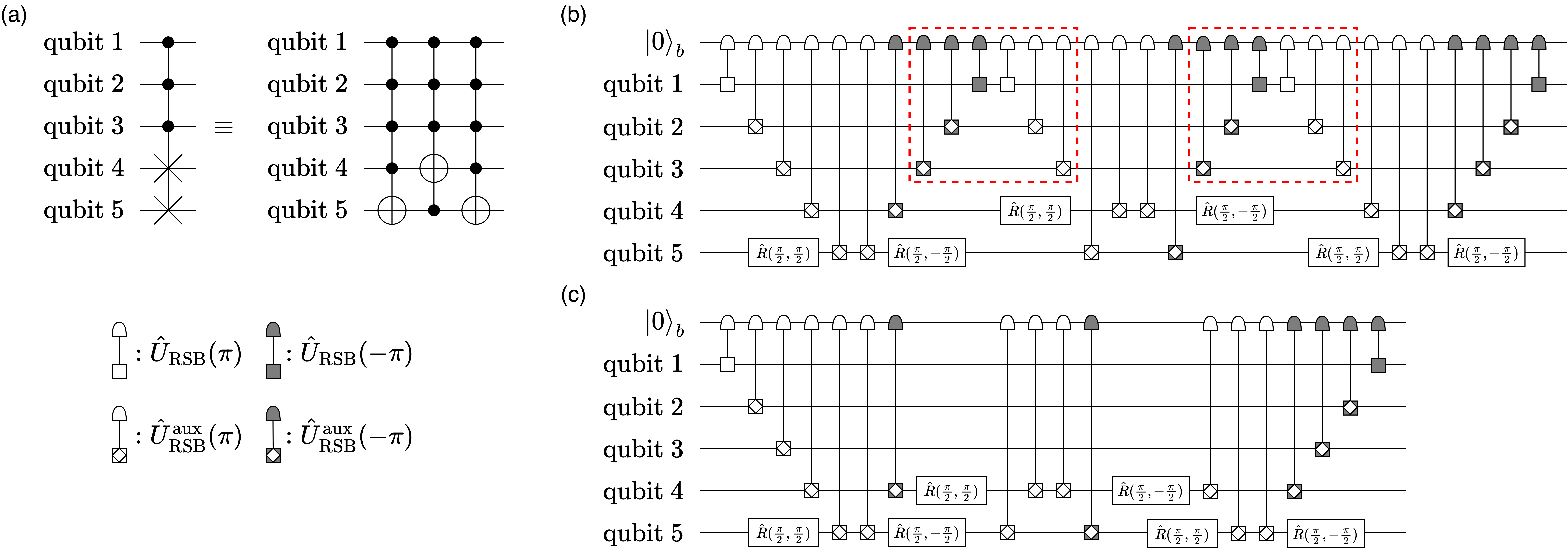}
    \caption{Implementation of \(N\)-CSWAP gate.
        (a) The \(N\)-CSWAP gate can be decomposed into three \((N+2)\)-Toffoli gates.
        (b) Implementation of three \((N+2)\)-Toffoli gates.
        \(4N\) RSB pulses inside the red dashed-line boxes can be eliminated by choosing an appropriate quantum circuit from Fig.~\ref{fig:toffoli}~(b--e) for each \((N+2)\)-Toffoli gate.
        (c) The resulting circuit for the \(N\)-CSWAP gate.}
    \label{fig:cswap}
\end{figure}

By providing flexibility in pulse sequence selection, the RSB gauge freedom allows for efficient implementation of successive multi-controlled gate operations.
Since the RSB \(\pi\) and \(-\pi\) pulses are inverses of each other, appropriate scheduling of adjacent multi-controlled gates enables the cancellation of redundant pulse sequences.
Specifically, when the second sequence of a preceding gate and the first sequence of a subsequent gate are inverses of each other, these adjacent sequences can be eliminated.
This cancellation reduces the total number of required RSB pulses, leading to a more efficient implementation of complex quantum circuits.
As a concrete example, consider the \(N\)-controlled SWAP (\(N\)-CSWAP) gate, whose standard decomposition into three consecutive Toffoli gates makes it a minimal testbed for pulse cancellation (Fig.~\ref{fig:cswap}~(a)).
We can assign the pulse configurations from Fig.~\ref{fig:toffoli}~(b--e) to the three Toffoli gates so that neighboring sequences form inverse pairs, allowing the intermediate RSB sequences to cancel directly (Fig.~\ref{fig:cswap}).

We now generalize the pulse cancellation to the successive implementation of \(M\) \(N\)-controlled gates sharing the same control and target qubits.
Here, \(\mathbf{x}_j\) (\(1 \leq j \leq M\)) denotes the binary control string of length \(N\) for the \(j\)th gate, ordered according to the sequence of RSB pulses, with \(\ket{g}\) and \(\ket{e}\) associated with \(0\) and \(1\), respectively.
Without exploiting the RSB gauge freedom---i.e., using only the circuit in Fig.~\ref{fig:mc_gate}~(b)---implementing \(M\) gates would require \(2M(N+1)\) RSB pulses.
In contrast, our method selects an appropriate variant circuit from Fig.~\ref{fig:mc_gate}~(b--e) for each gate, ensuring that the neighboring RSB sequences form inverse pairs at as many positions as possible. 
This pulse cancellation reduces the total number of required RSB pulses to
\begin{align}
    \label{eq:reduced_successive}
    2M(N+1) - 2\sum_{k=1}^{M-1}(c_k - 1),
\end{align}
where \(c_k\) is the first index at which the control strings \(\mathbf{x}_k\) and \(\mathbf{x}_{k+1}\) differ, and we set \(c_k = N+2\) if \(\mathbf{x}_k = \mathbf{x}_{k+1}\). 
The derivation of this expression and the detailed cancellation argument are provided in Appendix~\ref{app:deriv_of_successive}.

\begin{figure}
    \centering
    \includegraphics[width=1.0\linewidth]{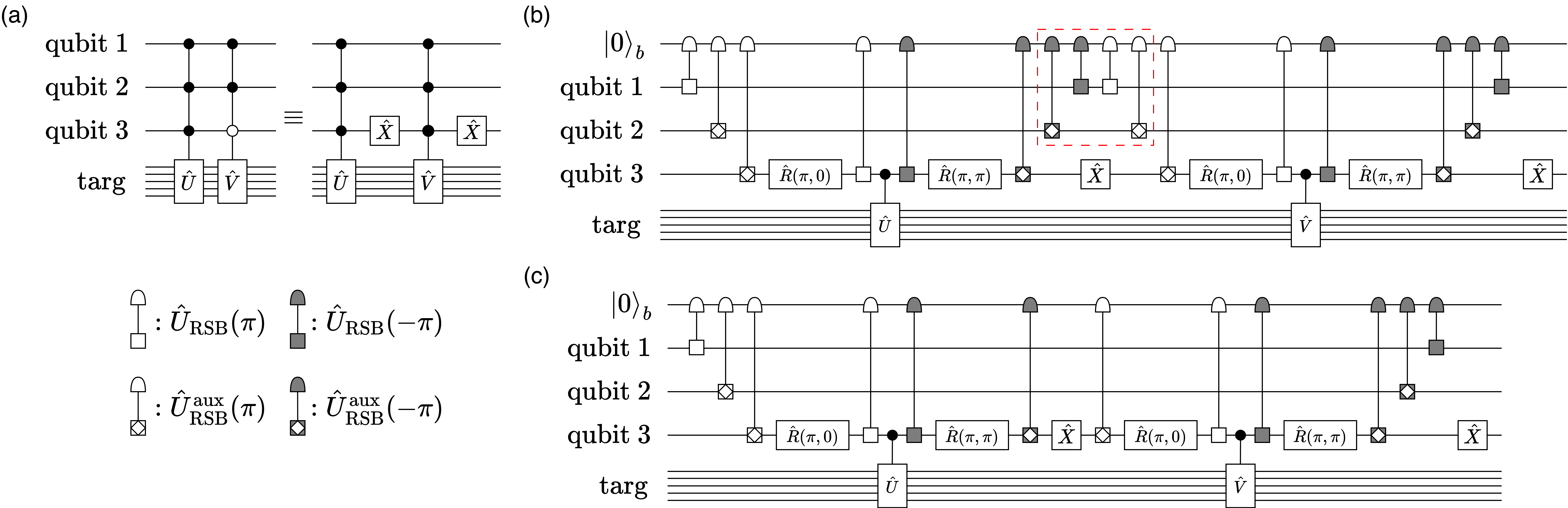}
    \caption{Implementation of two multi-controlled gates with different control conditions.
    (a) Circuit diagram of the two gates, which can be decomposed into multi-controlled gates with the same control condition and \(X\) gates.
    (b) The first and second multi-controlled gates are implemented using the quantum circuit shown in Fig.~\ref{fig:mc_gate}~(d). 
    The RSB pulses within the red dashed-line box can be eliminated.
    (c) The resulting circuit corresponding to (b).}
    \label{fig:two_controlled}
\end{figure}

To illustrate this cancellation, consider the simple case of two successive 3-controlled gates with control conditions \(\mathbf{x}_1 = 111\) and \(\mathbf{x}_2 = 110\) (Fig.~\ref{fig:two_controlled}~(a)).
Without cancellation, this requires \(2 \times 2 \times (3+1) = 16\) RSB pulses, and the first differing index is \(c_1 = 3\).
By applying our scheme, both gates are implemented using the variant shown in Fig.~\ref{fig:mc_gate}~(d).
As shown in Fig.~\ref{fig:two_controlled}~(b), the \(2(c_1 - 1) = 4\) RSB pulses within the red dashed box are directly eliminated.
As a result, the final circuit requires only 12 RSB pulses, as depicted in Fig.~\ref{fig:two_controlled}~(c).

\subsection{Numerical simulations of the 3-CSWAP gate}
\label{sec:numerical_validation}

To quantitatively evaluate our proposed pulse cancellation, we perform numerical simulations using the 3-CSWAP gate as a representative benchmark.
We decompose the 3-CSWAP gate into three 5-Toffoli gates (Fig.~\ref{fig:cswap}~(a)) and implement each Toffoli gate using the Cirac--Zoller scheme (Fig.~\ref{fig:toffoli}).
We compare two approaches: the standard approach, which does not use pulse cancellation (Fig.~\ref{fig:cswap}~(b)), and the proposed approach, which uses pulse cancellation (Fig.~\ref{fig:cswap}~(c)).

We simulate the open-system dynamics of the ion chain coupled to a shared motional mode by solving the Lindblad master equation using QuTiP~\cite{qutip5}.
The motional Hilbert space is truncated to a Fock cutoff of \(n_{\max}\,=\,10\), with the motional mode initialized in a thermal state with mean phonon number \(\bar{n} = 0.05\).
We include three error channels: motional heating (\(\hat{L}_h=\sqrt{\gamma_h}\,\hat{a}^\dagger\)), motional dephasing (\(\hat{L}_\phi=\sqrt{\gamma_\phi}\,\hat{a}^\dagger\hat{a}\)), and Zeeman dephasing (\(\hat{L}_z=\sqrt{\gamma_z}\,\ket{f}\bra{f}\)); all parameter values are given in Appendix~\ref{app:sim_params}.
We model RSB \(\pm\pi\) pulses as square pulses of duration \(t_{\mathrm{RSB}}=\pi/(\Omega\eta)\) and carrier pulses for \(\hat{R}(\pi/2,\pm\pi/2)\) as square pulses of duration \(t_{\text{carrier}}=\pi/2\Omega\).

We quantify the performance via the state fidelity 
\(F_{\ket{\psi_0}} = \text{Tr}(\hat{U}\ket{\psi_0}\bra{\psi_0}\hat{U}^{\dagger}\rho_\text{out})\), computed for each of the \(2^5=32\) input states \(\ket{\psi_0}\in\{\ket{g},\ket{e}\}^{\otimes5}\).
We report the mean fidelity over the two output states \(\ket{g\,xxxx}\) and \(\ket{e\,xxxx}\) separately, where each \(x \in \{g,e\}\), with error bars denoting the standard deviation across the 16 states within each group.

\begin{figure}[htb!]
    \centering
    \includegraphics[width=1.0\linewidth]{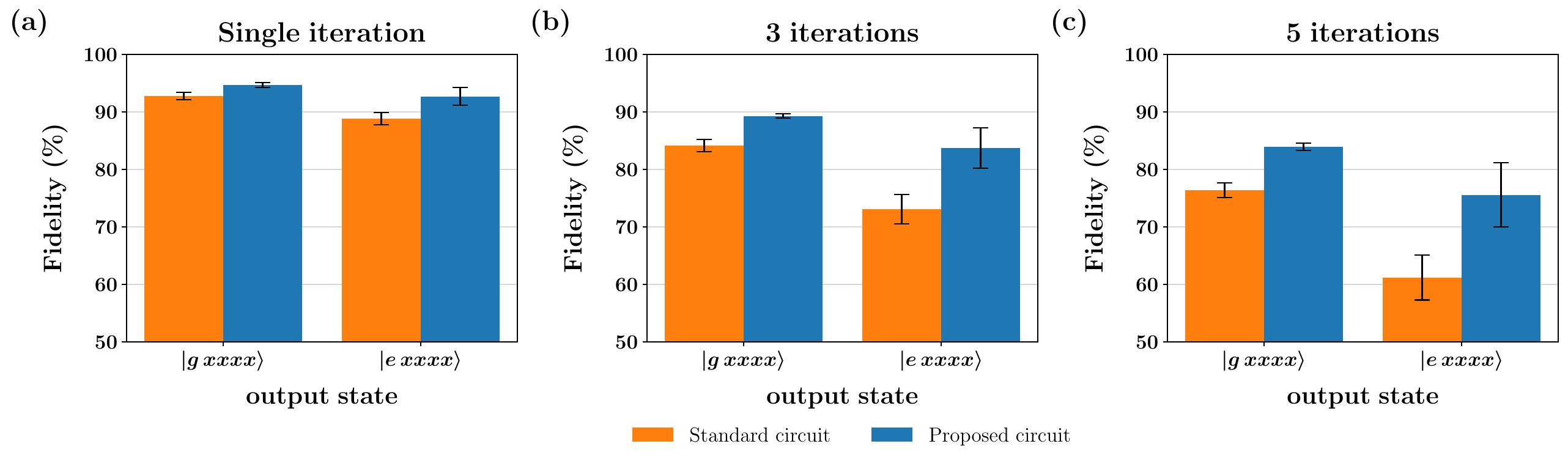}
    \caption{The state fidelities for the 3-controlled SWAP gate and its 3- and 5-iteration sequences, obtained from numerical simulations.
    The output states are grouped into \(\ket{g\,xxxx}\) and \(\ket{e\,xxxx}\), where each \(x \in \{g,e\}\); bars indicate the mean fidelity over each group and error bars denote the standard deviation.
    For each group, the mean fidelities of the standard circuit, constructed by concatenating three Toffoli gates using Fig.~\ref{fig:toffoli}~(b), and the proposed circuit (Fig.~\ref{fig:cswap}~(c)) are compared.
    (a) Single iteration.
    (b) 3 iterations.
    (c) 5 iterations.
    The simulation parameters and detailed state fidelities are provided in Appendix~\ref{app:sim_params}. }
    \label{fig:truth_table}
\end{figure}

Our results show that the proposed approach reduces the total gate time by \(39.6\,\%\) compared to the standard approach.
The average fidelity is enhanced from \(90.8\,\%\) in the standard approach to \(93.7\,\%\) in the proposed approach.
As shown in Fig.~\ref{fig:truth_table}~(a), the proposed approach improves the mean fidelity for both the \(\ket{g\,xxxx}\) and \(\ket{e\,xxxx}\) groups: the \(\ket{g\,xxxx}\) group achieves \(94.7 \pm 0.4\,\%\) (\(92.8 \pm 0.6\,\%\) for the standard approach), while the \(\ket{e\,xxxx}\) group achieves \(92.7 \pm 1.6\,\%\) (\(88.8 \pm 1.1\,\%\)), showing a particularly pronounced enhancement for the \(\ket{e\,xxxx}\) group.

The overall fidelity improvement can be attributed to the reduction in total gate time, which decreases the exposure to dissipation across all output states.
By examining the fidelity under each error channel individually---motional heating, motional dephasing, and Zeeman dephasing---we find that the fidelity difference for the \(\ket{e\,xxxx}\) group is primarily driven by motional heating.

When the first control qubit is in \(\ket{e}\), the gate sequence induces multiple Fock-state transitions in the motional mode (see Cases~3--6 in Appendix~\ref{appendix:detailed_v2}).
Because the higher motional Fock-state translates into a larger heating-induced error, the longer occupation time in \(\ket{1}_b\) for the \(\ket{e\,xxxx}\) group induces greater error accumulation.
Our pulse cancellation eliminates a number of RSB pulses, reducing the occupation time in \(\ket{1}_b\) and thereby suppressing the associated heating error.
This mechanism explains why the mean fidelity of the \(\ket{e\,xxxx}\) group is lower than that of the \(\ket{g\,xxxx}\) group, where no such transitions are required.

Moreover, to assess how the fidelity advantage scales with circuit depth, we perform 3 and 5 iterations of the 3-CSWAP gate under identical noise conditions.
As shown in Fig.~\ref{fig:truth_table}~(b) and~(c), the fidelity difference between the standard and proposed approaches increases with the number of iterations: the average fidelity of the proposed (standard) approach is \(86.5\,\%\) (\(78.6\,\%\)) for 3 iterations and \(79.7\,\%\) (\(68.8\,\%\)) for 5 iterations, confirming that the advantage of pulse cancellation becomes more pronounced as the circuit depth grows.

\section{Application to Linear Combination of Unitaries}
\label{sec: LCU}
\begin{figure}[htb!]
    \centering
    \begin{quantikz}
        \lstick[5]{\(\ket{g^{\otimes N}}\)} & \gate[wires=5, nwires=4]{\textsc{PREP}} & \octrl{1} \gategroup[6,steps=4,style={dashed, rounded corners, inner xsep=2pt}, background]{{SEL}} & \octrl{1} & \ \push{\ldots} \ & \ctrl{1} & \gate[wires=5, nwires=4]{\textsc{PREP}^\dagger} & \qw &  \\
        & & \octrl{1} & \octrl{1} & \ \push{\ldots} \ & \ctrl{1} & & \qw \\
        & & \octrl{2} & \octrl{2} & \ \push{\ldots} \ & \ctrl{2} & & \qw \\
        \ \vdots \ && \gate[nwires=1,style={fill=white,draw=none}]{\vdots} & \gate[nwires=1,style={fill=white,draw=none}]{\vdots} & \gate[nwires=1,style={fill=white,draw=none}]{\ddots}  & \gate[nwires=1,style={fill=white,draw=none}]{\vdots} & & \ \vdots \ \\
        & & \octrl{1} & \ctrl{1} & \ \push{\ldots} \ &  \ctrl{1} & & \qw \\
        \lstick{\(\mathrm{targ}\)} & \qwbundle[alternate]{} & \gate{U_0} \qwbundle[alternate]{} & \gate{U_1} \qwbundle[alternate]{} &\qwbundle[alternate]{} \ \ldots\ & \gate{U_{L-1}} \qwbundle[alternate]{} & \qwbundle[alternate]{} & \qwbundle[alternate]{}
    \end{quantikz}
    \caption{Quantum circuit implementing the block encoding of Eq.~\eqref{eqn: LCU} using the LCU method, where \(\widehat{\mathrm{PREP}}\) and \(\widehat{\mathrm{SEL}}\) are defined in Eqs.~\eqref{eqn: PREP} and \eqref{eqn: SEL}, respectively~\cite{childs2012hamiltonian,Kane2025blockencodingbosons}.
    Here, \(\ket{g^{\otimes N}}\) and \(\mathrm{targ}\) represent ancillary and target qubits.}
    \label{fig: LCU circuit}
\end{figure}
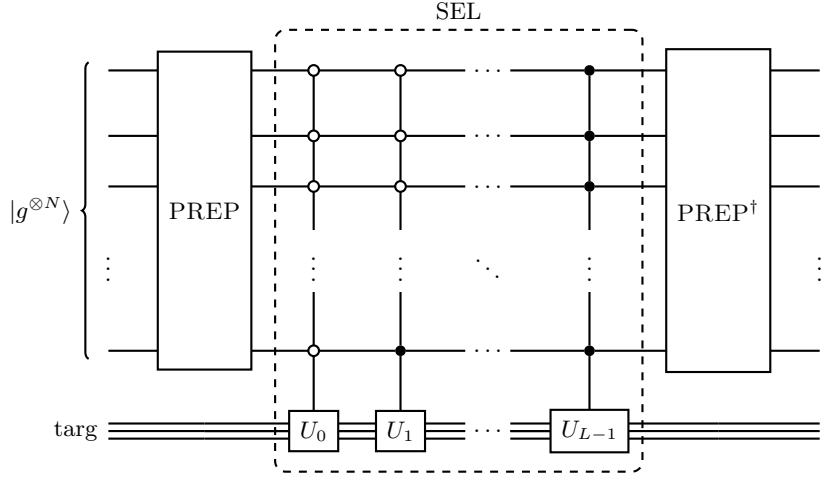

One of the key applications of our work is the implementation of the LCU method~\cite{childs2012hamiltonian}, a prominent method for constructing block encodings~\cite{Low2019hamiltonian,camps2023explicitquantumcircuitsblock,Kane2025blockencodingbosons} of non-unitary operators.
In the LCU framework, a non-unitary operator \(\hat{A}\) is expressed as a linear combination of unitaries:
\begin{align}
    \label{eqn: LCU}
    \hat{A} = \sum_{l=0}^{L-1}a_l\hat{U}_l,
\end{align}
where each \(\hat{U}_l\) is a unitary operator and \(a_l\) is a positive real coefficient.
The LCU circuit employs \(N = \lceil\log_{2}{L}\rceil\) ancillary qubits and consists of a state preparation operator \(\widehat{\mathrm{PREP}}\), its adjoint \(\widehat{\mathrm{PREP}}^{\dagger}\), and a select operator \(\widehat{\mathrm{SEL}}\).
The state preparation operator encodes the coefficients \(a_l\) into the amplitudes of the ancillary state, and the select operator conditionally applies each \(\hat{U}_l\) based on the ancillary state \(\ket{l}\), where \(\ket{l}\) is the computational basis state of \(N\) ancillary qubits in the \(\{\ket{g},\ket{e}\}\) basis, with \(\ket{g}\) and \(\ket{e}\) associated with \(0\) and \(1\), respectively.
These operators are defined, respectively, as
\begin{gather}
    \label{eqn: PREP}
    \widehat{\mathrm{PREP}}\ket{g^{\otimes N}}=\sum_{l=0}^{L-1}\sqrt{\frac{a_{l}}{s}}\ket{l},\\
    \label{eqn: SEL}
    \widehat{\mathrm{SEL}}=\sum_{l=0}^{L-1}\ket{l}\bra{l}\otimes\hat{U}_{l},
\end{gather}
where the normalization factor is \(s = \sum_{l=0}^{L-1} \lvert a_l \rvert\).
The LCU circuit operator is given by \((\widehat{\mathrm{PREP}}^{\dagger}\otimes\hat{I})\,\widehat{\mathrm{SEL}}\,(\widehat{\mathrm{PREP}}\otimes \hat{I})\), and it forms a block encoding of \(\hat{A}\) (Fig.~\ref{fig: LCU circuit}):
\begin{align}
    \label{eqn: LCU_BE}
    (\bra{g^{\otimes N}}\otimes\hat{I})\;(\widehat{\mathrm{PREP}}^{\dagger}\otimes\hat{I})\;\widehat{\mathrm{SEL}}\;(\widehat{\mathrm{PREP}}\otimes\hat{I})\;(\ket{g^{\otimes N}}\otimes\hat{I}) = \frac{\hat{A}}{s}.
\end{align}
In general, the primary resource cost comes from the select operator, which consists of \(L\) successive \(N\)-controlled gates.

As discussed in Sec.~\ref{sec:main_result}, given access to single-controlled gates, \(\widehat{\mathrm{SEL}}\) can be implemented with \(2L(N+1) = 2L(\lceil\log_2L\rceil+1)\) RSB pulses without exploiting the RSB gauge freedom, and the number of RSB pulses can be reduced by applying pulse cancellation.
For every \(0 \leq l \leq L -1\), we denote the control condition on the ancillary qubits for \(\hat{U}_l\) as the \(N\)-bit string \(\mathbf{x}_l^{(N)} = x_{N-1}x_{N-2}\cdots x_1x_0\), where \(l = \sum_{j=0}^{N-1} x_j 2^j\).
While the exact number of RSB pulses for an arbitrary \(L\) is hard to determine, we show that for the \(L = 2^N\) case, the total number of RSB pulses becomes \(6L-4\) (the derivation is detailed in Appendix~\ref{app:deriv_LCU}).
As shown in Fig.~\ref{fig: comparison}, for arbitrary \(L\), the number of RSB pulses is approximately \(6L-4\). Therefore, the number of RSB pulses reduces from \(\mathcal{O}(L\log L)\) to \(\mathcal{O}(L)\), and the ratio \(N_{\mathrm{RSB}}^{\mathrm{canc}}/N_{\mathrm{RSB}}\) decreases with increasing \(L\).

\begin{figure}[htb!]
    \centering
    \includegraphics[width=0.6\linewidth]{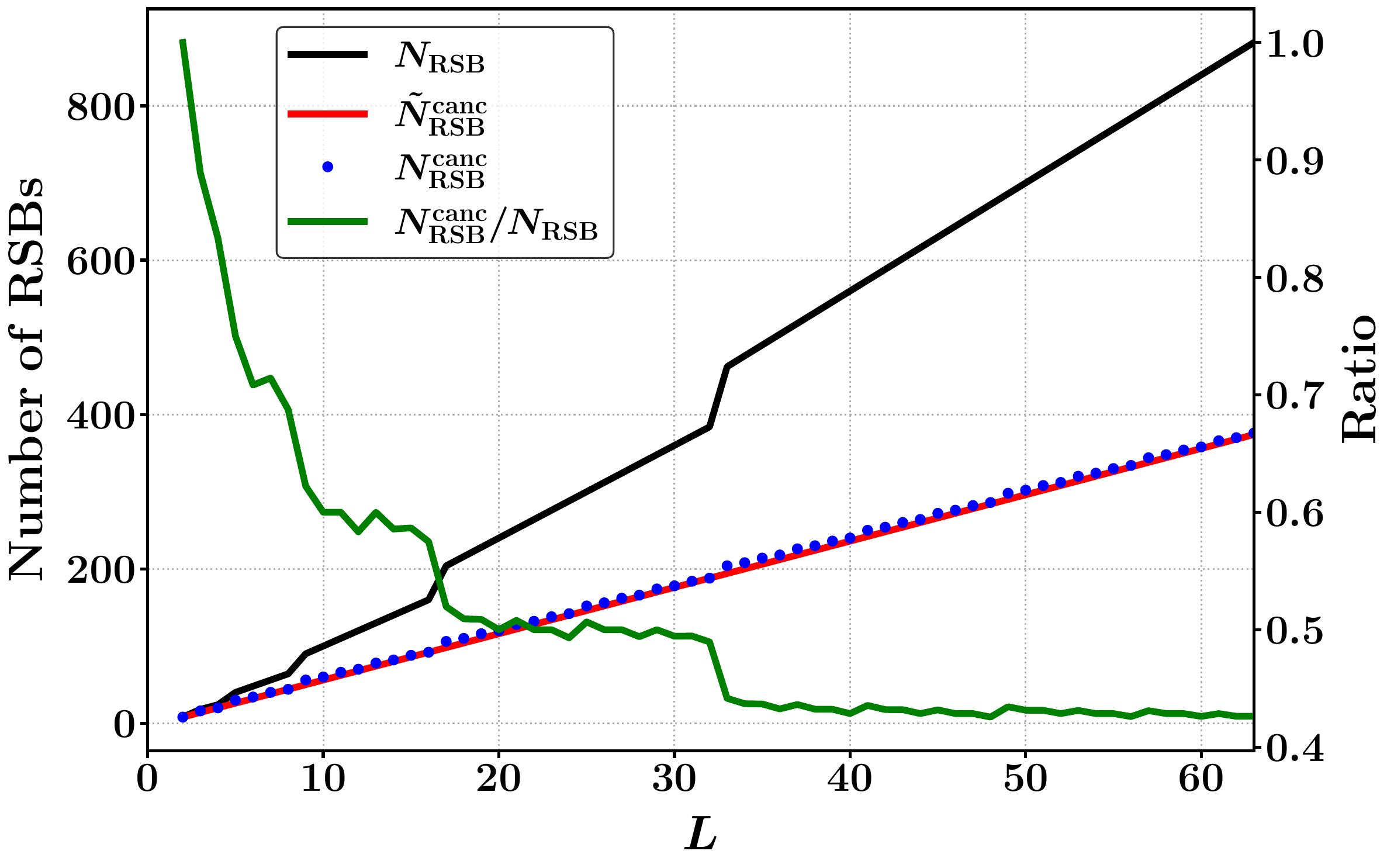}
    \caption{The number of RSB pulses and the relative ratio plotted against \(L\), the number of unitaries in the linear combination.
    The black line and the blue dots represent the RSB count for different values of \(L\) before and after cancellation, \(N_{\mathrm{RSB}}\) and \(N^{\mathrm{canc}}_{\mathrm{RSB}}\).
    The red line shows the RSB count after cancellation obtained from \(L=2^{N}\), \(\tilde{N}_{\mathrm{RSB}}^{\mathrm{canc}}=6L-4\).
    The green line indicates the relative ratio of RSB pulses after cancellation, \(N^{\mathrm{canc}}_{\mathrm{RSB}}/{N_{\mathrm{RSB}}}\).
    }
    \label{fig: comparison}
\end{figure}

\begin{figure}[htb!]
    \centering
    \includegraphics[width=1.0\linewidth]{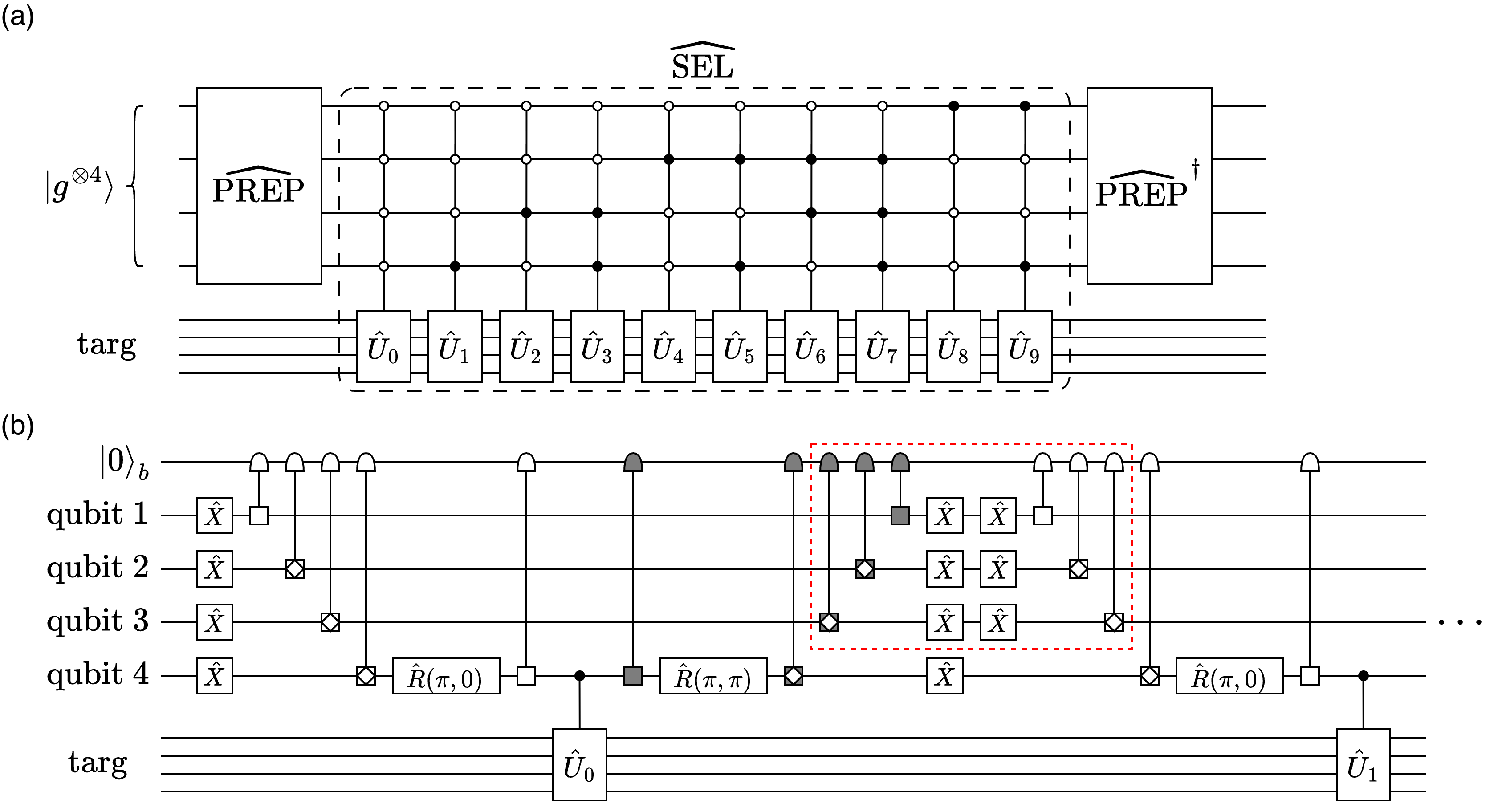}
    \caption{(a) The LCU circuit for a block encoding of \(\hat{H}\), as defined in Eq.~\eqref{eqn: LCU ex}.
        It requires four ancillary qubits.
        (b) The implementation of the select operator using pulse cancellation.
        The RSB pulses inside the red dashed-line boxes are canceled.}
    \label{fig: circuit ex}
\end{figure}

We present a simple example in which the Hamiltonian \(\hat{H}\) is expressed as a linear combination of ten unitaries:
\begin{align}
    \label{eqn: LCU ex}
    \hat{H} = \sum_{l=0}^{9} a_l \hat{U}_l.
\end{align}
The corresponding LCU circuit for a block encoding of \(\hat{H}\) is illustrated in Fig.~\ref{fig: circuit ex}.
It requires \(\lceil\log_2 10\rceil = 4\) ancillary qubits, and its select operator involves ten 4-controlled gates.
Without pulse cancellation, each 4-controlled gate requires 10 RSB pulses, resulting in a total of 100 RSB pulses.
In the proposed method, 40 of these pulses---those highlighted by the red dashed boxes in Fig.~\ref{fig: circuit ex}(b)---cancel, reducing the total to 60 pulses.

\begin{figure}
    \centering
    \includegraphics[width=0.6\linewidth]{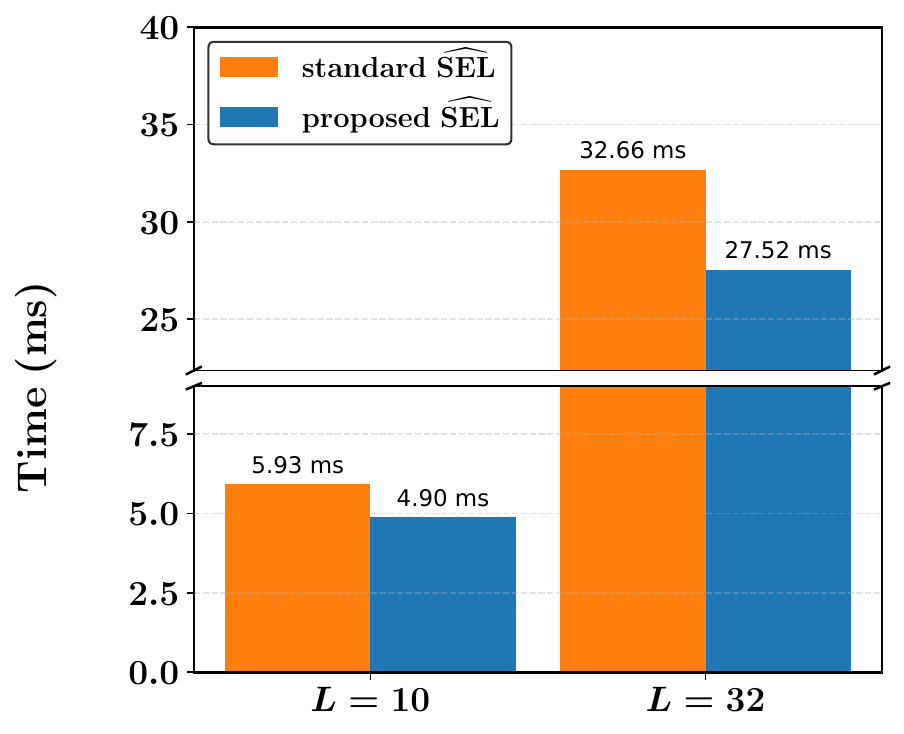}
    \caption{Comparison of the total gate time for the standard \(\widehat{\mathrm{SEL}}\) (orange) and proposed \(\widehat{\mathrm{SEL}}\) (blue) implementations for \(L=10\) and \(L=32\), in \(\mathrm{ms}\).
    The standard implementation constructs each multi-controlled gate using the circuit in Fig.~\ref{fig:mc_gate}~(b), while the proposed implementation exploits pulse cancellation by selecting appropriate variants from Fig.~\ref{fig:mc_gate}~(b--e).}
    \label{fig:lcu gate time}
\end{figure}

To assess the practical impact of the pulse reduction, we compare the total execution time of the standard and proposed \(\widehat{\mathrm{SEL}}\) implementations for an LCU decomposition of a Pauli Hamiltonian.
We consider a Hamiltonian given by an equal-weight linear combination of all Pauli strings acting on \(n\) target qubits.
These Pauli strings are grouped into \(L\) sets of mutually commuting terms, with \(L=10\) for \(n=2\) and \(L=32\) for \(n=3\) as concrete examples.
For gate-time estimates, we adopt the same simulation parameters as in Sec.~\ref{sec:numerical_validation}.
Here, \(t_{\mathrm{RSB}}\) denotes the gate time of an RSB \(\pi\) pulse, the gate time of a single controlled-Pauli gate is set to \(4\,t_{\mathrm{RSB}}\), and that of a controlled \(m\)-fold tensor product of Pauli operators to \((2m-1) \times 4\,t_{\mathrm{RSB}}\).
As shown in Fig.~\ref{fig:lcu gate time}, the proposed approach reduces the total gate time from \(5.93\,\mathrm{ms}\) to \(4.90\,\mathrm{ms}\) for \(L=10\), and from \(32.66\,\mathrm{ms}\) to \(27.52\,\mathrm{ms}\) for \(L=32\), demonstrating that the execution time savings grow with increasing \(L\).

\section{Conclusion and discussion}
\label{sec: conclusion}

In this paper, we introduced a pulse-level framework for multi-controlled gate implementation in the Cirac--Zoller scheme, exploiting the freedom in the choice of RSB \(\pm\pi\) pulses as a gauge freedom.
This freedom enables a pulse cancellation that reduces the RSB pulse count in successive multi-controlled gates without modifying the logical operation.

Applying this pulse cancellation to the 3-controlled SWAP gate, our numerical simulations under realistic noise confirmed a \(39.6\%\) reduction in gate time and an improvement in average fidelity from \(90.8\%\) to \(93.7\%\), with the motional-sensitive \(\ket{e\,xxxx}\) group showing the largest gain (\(88.8\pm1.1\%\!\to\!92.7\pm1.6\%\)).
As a broader application, we derived a general formula for the RSB pulse count of successive multi-controlled gates, which we used to show that the select operator of an LCU block encoding over \(L\) unitaries can be implemented with \(\mathcal{O}(L)\) RSB pulses instead of \(\mathcal{O}(L\log L)\)---a reduction that translates to \(17\,\%\) and \(16\,\%\) gate-time savings at \(L=10\) and \(L=32\), respectively.

Our approach inherits both the strengths and limitations of the Cirac--Zoller platform. The recent realization of a 5-Toffoli gate in \({}^{171}\mathrm{Yb}^{+}\) ions~\cite{fang2023realization} demonstrates the feasibility of this platform, while limited auxiliary-state coherence and sensitivity to motional heating remain practical concerns.
Scaling the framework to longer pulse sequences or larger ion chains, where these challenges intensify, remains an open question.
Nonetheless, the optimizations developed here reduce precisely the pulse overhead most susceptible to such noise, suggesting a viable path for near-term trapped-ion hardware.

Two complementary directions deserve further exploration.
First, we expect that the proposed reduction scheme could be extended to alternative gate implementations.
Our current scheme employs only RSB transitions, including those on the auxiliary state; however, since blue-sideband (BSB) transitions~\cite{Leibfried2003quantum} share an analogous structure with RSB transitions, we expect that the same gauge freedom in the choice of pulse signs extends to BSB-based implementations.
Furthermore, combining RSB and BSB transitions---including those on the auxiliary state---could enable a M\o lmer--S\o rensen (MS)-gate-based implementation~\cite{PhysRevLett.82.1835,PhysRevLett.82.1971} that inherits both our pulse savings and the intrinsic robustness of MS gates against motional thermal fluctuations.
Second, we anticipate that the core idea of exploiting gauge freedom is not specific to the Cirac--Zoller scheme.
More broadly, our work illustrates how physical-layer symmetries---often hidden under gate-level abstractions---can be exploited to reduce quantum circuit costs.
Identifying and leveraging such symmetries may prove a general strategy for bridging abstract gate counts and hardware-level pulse budgets as quantum processors scale.

\section*{Acknowledgments}
\label{sec: ack}
The authors thank Junki Kim and Taeyoung Choi for in-depth discussion. 
This work received partial support from multiple sources:
[1] Basic Science Research Program through the National Research Foundation of Korea (NRF), funded by the Ministry of Science and ICT (RS-2023-NR068116, RS-2025-03532992).
[2] Institute for Information \& Communications Technology Promotion (IITP) grant funded by the Korea government (MSIP) (No. 2019-0-00003), which focuses on the research and development of core technologies for programming, running, implementing, and validating fault-tolerant quantum computing systems.
[3] Korean ARPA-H Project via the Korea Health Industry Development Institute (KHIDI); Ministry of Health and Welfare, Republic of Korea (RS-2025-25456722).
[4] Yonsei University Research Fund under project number 2025-22-0140. 

\section*{Appendix}
\appendix

\section{Detailed description of the multi-controlled gate}
\label{appendix:detailed_v2}

\begin{figure}[htb!]
    \centering
    \includegraphics[width=1.0\linewidth]{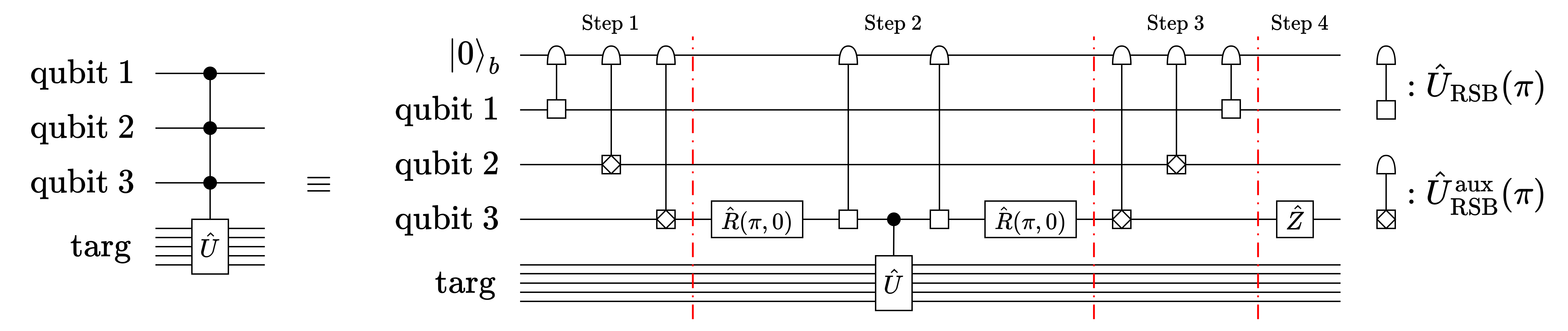}
    \caption{Quantum circuit of Fig.~\ref{fig:mc_gate}~(b) with vertical dashed lines separating the circuit into segments corresponding to Steps~1--4.}
    \label{fig:mc_circuit_dynamics}
\end{figure}

In this section, we provide a detailed description of how the quantum circuits shown in Fig.~\ref{fig:mc_gate}~(b--e) implement a multi-controlled gate.
These circuits can be understood in terms of how the control condition is encoded, applied, and decoded.
For simplicity, we focus on the case of three control qubits; it is sufficient to capture the circuit behavior for general \(N\).
Subscripts \(j\) and '\(\text{targ}\)' label the \(j\)th control qubit and the target register, respectively.

The circuit behavior can be fully characterized by analyzing the following six cases:
\begin{itemize}
    \item Case 1: The first and last control qubits are in \(\ket{g}\).
    Example: \(\ket{g}_{1}\ket{e}_{2}\ket{g}_{3}\).
    \item Case 2: The first control qubit is in \(\ket{g}\) and the last control qubit is in \(\ket{e}\).
    Example: \(\ket{g}_{1}\ket{e}_{2}\ket{e}_{3}\).

    \item Case 3: The first control qubit is in \(\ket{e}\), the last control qubit is in \(\ket{g}\), and there is at least one of the remaining control qubits in \(\ket{g}\).
    Example: \(\ket{e}_1\ket{g}_2\ket{g}_3\).

    \item Case 4: The first and the last control qubits are in \(\ket{e}\), and there is at least one of the remaining control qubits in \(\ket{g}\).
    Example: \(\ket{e}_1\ket{g}_2\ket{e}_3\).

    \item Case 5: All control qubits are in \(\ket{e}\) except the last one in \(\ket{g}\).
    Example: \(\ket{e}_1\ket{e}_2\ket{g}_3\).

    \item Case 6: All control qubits are in \(\ket{e}\).
    Example: \(\ket{e}_{1}\ket{e}_{2}\ket{e}_{3}\).
\end{itemize}

The circuit proceeds in four steps, as shown in Fig.~\ref{fig:mc_circuit_dynamics}.
In each step, we analyze the dynamics for the six cases above.
We denote the RSB pulses in the first sequence by RSB \((-1)^{s_1}\pi\) pulses and those in the second sequence by RSB \((-1)^{s_2}\pi\) pulses, with \(s_1,s_2 \in \{0,1\}\).

\begin{itemize}
    \item \textbf{Step 1: Encoding of the control condition.} 
    We apply \(N\) RSB \((-1)^{s_1}\pi\) pulses from the first to the last control qubit:
    \(\hat{U}_{\mathrm{RSB}}((-1)^{s_1}\pi)\) on the first control qubit and \(\hat{U}_{\mathrm{RSB}}^{\mathrm{aux}}((-1)^{s_1}\pi)\) on the remaining ones.
    When all control qubits are in \(\ket{e}\), the control condition is satisfied and exactly one phonon is loaded on the motional mode; otherwise, the motional mode remains in \(\ket{0}_b\).
    \begin{itemize}
        \item \textbf{Case 1 \& 2.} Since the first control qubit is in \(\ket{g}\), no energy exchange occurs:
        \begin{align}
            &\text{Case 1: }\ket{0}_b\ket{g}_{1}\ket{e}_{2}\ket{g}_{3}\ket{\psi}_{\mathrm{targ}}
            \rightarrow
            \ket{0}_b\ket{g}_{1}\ket{e}_{2}\ket{g}_{3}\ket{\psi}_{\mathrm{targ}}, \\
            &\text{Case 2: }\ket{0}_b\ket{g}_{1}\ket{e}_{2}\ket{e}_{3}\ket{\psi}_{\mathrm{targ}}
            \rightarrow
            \ket{0}_b\ket{g}_{1}\ket{e}_{2}\ket{e}_{3}\ket{\psi}_{\mathrm{targ}}.
        \end{align}
        \item \textbf{Case 3--5.}
        The first control qubit is in \(\ket{e}\), so the energy exchange occurs by the first RSB pulse. 
        However, since at least one of the remaining control qubits is in \(\ket{g}\), the phonon is absorbed, leaving the bosonic mode in \(\ket{0}_b\).
        The accumulated phase is \((-\mathrm{i})^2(-1)^{2s_1}=-1\):
        \begin{align}
            &\text{Case 3: }\ket{0}_b\ket{e}_{1}\ket{g}_{2}\ket{g}_{3}\ket{\psi}_{\mathrm{targ}}
            \rightarrow
            (-1)\ket{0}_b\ket{g}_{1}\ket{f}_{2}\ket{g}_{3}\ket{\psi}_{\mathrm{targ}},\\
            &\text{Case 4: }\ket{0}_b\ket{e}_{1}\ket{g}_{2}\ket{e}_{3}\ket{\psi}_{\mathrm{targ}}
            \rightarrow
            (-1)\ket{0}_b\ket{g}_{1}\ket{f}_{2}\ket{e}_{3}\ket{\psi}_{\mathrm{targ}},\\
            &\text{Case 5: }\ket{0}_b\ket{e}_{1}\ket{e}_{2}\ket{g}_{3}\ket{\psi}_{\mathrm{targ}}
            \rightarrow
            (-1)\ket{0}_b\ket{g}_{1}\ket{e}_{2}\ket{f}_{3}\ket{\psi}_{\mathrm{targ}}.
        \end{align}
    
        \item \textbf{Case 6.}
        When all control qubits are in \(\ket{e}\), only the first RSB pulse exchanges energy with the motional mode.
        The accumulated phase is \(-\mathrm{i}(-1)^{s_1}\):
        \begin{align}
            \ket{0}_b\ket{e}_{1}\ket{e}_{2}\ket{e}_{3}\ket{\psi}_{\mathrm{targ}}
            \rightarrow
            -\mathrm{i}(-1)^{s_1}\ket{1}_b\ket{g}_{1}\ket{e}_{2}\ket{e}_{3}\ket{\psi}_{\mathrm{targ}}.
        \end{align}
    \end{itemize}
    
    \item \textbf{Step 2: Conditional operation on the target register.}  
    By applying an \(\hat{R}(\pi,s_1\pi)\) gate to the last control qubit followed by an RSB \((-1)^{s_1}\pi\) pulse, the last control qubit is excited to \(\ket{e}\) when the motional mode is in the state \(\ket{1}_b\).
    For the other cases, the last control qubit remains in \(\ket{g}\) or \(\ket{f}\).
    As a result, the last control qubit encodes the control condition and is used to apply the conditional operation \(\hat{U}\) on the target register.
    The last control qubit state is restored by applying an RSB \((-1)^{s_2}\pi\) pulse followed by an \(\hat{R}(\pi,s_2\pi)\) gate.
    
    \begin{itemize}
        \item \textbf{Case 1 \& 3.}  
        Since the motional mode is in \(\ket{0}_b\) and the third control qubit is in \(\ket{g}\), the first rotation gate and the subsequent RSB pulse remain the third control qubit in \(\ket{g}\).
        Therefore, the conditional operator \(\hat{U}\) is not applied.
        The accumulated phase is \(((-1)^{s_1}(-\mathrm{i}))^2 ((-1)^{s_2}(-\mathrm{i}))^2 = 1\):
        \begin{align}
            &\text{Case 1: }\ket{0}_b\ket{g}_{1}\ket{e}_{2}\ket{g}_{3}\ket{\psi}_{\mathrm{targ}}
            \rightarrow
            \ket{0}_b\ket{g}_{1}\ket{e}_{2}\ket{g}_{3}\ket{\psi}_{\mathrm{targ}},\\
            &\text{Case 3: }(-1)\ket{0}_b\ket{g}_{1}\ket{f}_{2}\ket{g}_{3}\ket{\psi}_{\mathrm{targ}}
            \rightarrow
            (-1)\ket{0}_b\ket{g}_{1}\ket{f}_{2}\ket{g}_{3}\ket{\psi}_{\mathrm{targ}}.
        \end{align}
    
        \item \textbf{Case 2 \& 4.}  
        Since the motional mode is in \(\ket{0}_b\) and the third control qubit is in \(\ket{e}\), the first rotation gate changes the third control qubit to \(\ket{g}\), and the subsequent RSB pulses thus do not exchange energy.
        Therefore, the conditional operator \(\hat{U}\) is not applied. The accumulated phase is \((-\mathrm{i})^{s_1}(-\mathrm{i})(-\mathrm{i})^{s_2}(-\mathrm{i}) = (-1)^{s_1+s_2+1}\):
        \begin{align}
            &\text{Case 2: }\ket{0}_b\ket{g}_{1}\ket{e}_{2}\ket{e}_{3}\ket{\psi}_{\mathrm{targ}}
            \rightarrow
            (-1)^{s_1+s_2+1}\ket{0}_b\ket{g}_{1}\ket{e}_{2}\ket{e}_{3}\ket{\psi}_{\mathrm{targ}},
            \\
            &\text{Case 4: }(-1)\ket{0}_b\ket{g}_{1}\ket{f}_{2}\ket{e}_{3}\ket{\psi}_{\mathrm{targ}}
            \rightarrow
            (-1)^{s_1+s_2}\ket{0}_b\ket{g}_{1}\ket{f}_{2}\ket{e}_{3}\ket{\psi}_{\mathrm{targ}}.
        \end{align}

        \item \textbf{Case 5.} Since the third control qubit is in \(\ket{f}\), all RSB pulses and single-qubit rotations do not act on the third control qubit.
        Therefore, the conditional operator \(\hat{U}\) is not applied. There is no accumulated phase:
        \begin{align}
            (-1)\ket{0}_b\ket{g}_1\ket{e}_2\ket{f}_3\ket{\psi}_\mathrm{targ} \rightarrow (-1)\ket{0}_b\ket{g}_1\ket{e}_2\ket{f}_3\ket{\psi}_\mathrm{targ}.
        \end{align}
        
        \item \textbf{Case 6.}  
        Since the motional mode is in \(\ket{1}_b\) and the third control qubit is in \(\ket{e}\), the first rotation gate changes the third control qubit to \(\ket{g}\), and then the subsequent RSB pulse exchanges energy with the motional mode, returning the qubit to \(\ket{e}\) while absorbing the phonon.
        Therefore, the conditional operator \(\hat{U}\) is applied. The accumulated phase is \(((-1)^{s_1}(-\mathrm{i}))^2 ((-1)^{s_2}(-\mathrm{i}))^2 = 1\):
        \begin{align}
            -\mathrm{i}(-1)^{s_1}\ket{1}_b\ket{g}_{1}\ket{e}_{2}\ket{e}_{3}\ket{\psi}_{\mathrm{targ}}
            \rightarrow
            -\mathrm{i}(-1)^{s_1}\ket{1}_b\ket{g}_{1}\ket{e}_{2}\ket{e}_{3}\hat{U}\ket{\psi}_{\mathrm{targ}}.
        \end{align}
    \end{itemize}
    
    \item \textbf{Step 3: Decoding the control condition.} 
    We apply \(N\) RSB \((-1)^{s_2}\pi\) pulses from the last to the first control qubit: \(\hat{U}_{\mathrm{RSB}}((-1)^{s_2}\pi)\) on the first control qubit and \(\hat{U}_{\mathrm{RSB}}^{\mathrm{aux}}((-1)^{s_2}\pi)\) on the remaining ones.
    All control qubits and the motional mode are restored, giving the multi-controlled gate up to relative phase, which can be corrected by a single-qubit rotation gate.
    
    \begin{itemize}
        \item \textbf{Case 1 \& 2.}
        No energy exchange occurs:
        \begin{align}
           &\text{Case 1: }\ket{0}_b\ket{g}_{1}\ket{e}_{2}\ket{g}_{3}\ket{\psi}_{\mathrm{targ}}
            \rightarrow
            \ket{0}_b\ket{g}_{1}\ket{e}_{2}\ket{g}_{3}\ket{\psi}_{\mathrm{targ}}, \\
            &\text{Case 2: }(-1)^{s_1+s_2+1}\ket{0}_b\ket{g}_{1}\ket{e}_{2}\ket{e}_{3}\ket{\psi}_{\mathrm{targ}}
            \rightarrow
            (-1)^{s_1+s_2+1}\ket{0}_b\ket{g}_{1}\ket{e}_{2}\ket{e}_{3}\ket{\psi}_{\mathrm{targ}}.
        \end{align}
        
        \item \textbf{Case 3--5.}  
        There are two energy exchanges, contributing to an accumulated phase of \(((-\mathrm{i})(-1)^{s_2})^2=-1\):
        \begin{align}
            &\text{Case 3: }(-1)\ket{0}_b\ket{g}_{1}\ket{f}_{2}\ket{g}_{3}\ket{\psi}_{\mathrm{targ}} \rightarrow \ket{0}_b\ket{e}_1\ket{g}_2\ket{g}_3\ket{\psi}_\mathrm{targ},\\
            &\text{Case 4: } (-1)^{s_1+s_2}\ket{0}_b\ket{g}_1\ket{f}_2\ket{e}\ket{\psi}_\mathrm{targ} \rightarrow (-1)^{s_1+s_2+1}\ket{0}_b\ket{e}_1\ket{g}_2\ket{e}\ket{\psi}_\mathrm{targ},\\
            &\text{Case 5: } (-1)\ket{0}_b\ket{g}_1\ket{e}_2\ket{f}_3\ket{\psi}_\mathrm{targ} \rightarrow \ket{0}_b\ket{e}_1\ket{e}_2\ket{g}_3\ket{\psi}_\mathrm{targ}.
        \end{align}
    
        \item \textbf{Case 6.}  
        There is one energy exchange, contributing to an accumulated phase of \((-\mathrm{i})(-1)^{s_2}\):       \begin{align}
            -\mathrm{i}(-1)^{s_1}\ket{1}_b\ket{g}_{1}\ket{e}_{2}\ket{e}_{3}\hat{U}\ket{\psi}_{\mathrm{targ}}
            \rightarrow
            (-1)^{s_1+s_2+1}\ket{0}_b\ket{e}_{1}\ket{e}_{2}\ket{e}_{3}\hat{U}\ket{\psi}_{\mathrm{targ}}.
        \end{align}
    \end{itemize}
    
    \item \textbf{Step 4: Phase correction.} 
    The phase for each case depends on the last control qubit state: \(1\) for \(\ket{g}\) and \((-1)^{s_1+s_2+1}\) for \(\ket{e}\).
    These phases can be corrected by applying \(\hat{Z}^{s_1+s_2 + 1}\) on the last control qubit.
    Since \(\hat{Z}^{2} = \hat{I}\), \(\hat{Z}\) is only applied if \(s_1 = s_2\) (see Fig.~\ref{fig:mc_gate}):
    
    \begin{itemize}
        \item \textbf{Case 1 \& 3 \& 5.}  
        The third control qubit is in \(\ket{g}\), so the \(Z\) gate has no effect in this case:
        \begin{align}
            &\text{Case 1: } \ket{0}_b\ket{g}_1\ket{e}_2\ket{g}_3\ket{\psi}_\mathrm{targ} \rightarrow \ket{0}_b\ket{g}_1\ket{e}_2\ket{g}_3\ket{\psi}_\mathrm{targ},\\
            &\text{Case 3: } \ket{0}_b\ket{e}_1\ket{g}_2\ket{g}_3\ket{\psi}_\mathrm{targ} \rightarrow \ket{0}_b\ket{e}_1\ket{g}_2\ket{g}_3\ket{\psi}_\mathrm{targ},\\
            &\text{Case 5: } \ket{0}_b\ket{e}_1\ket{e}_2\ket{g}_3\ket{\psi}_\mathrm{targ} \rightarrow \ket{0}_b\ket{e}_1\ket{e}_2\ket{g}_3\ket{\psi}_\mathrm{targ}.
        \end{align}
    
        \item \textbf{Case 2 \& 4 \& 6.}  
        By applying \(\hat{Z}^{s_1+s_2+1}\) to the third control qubit, the accumulated relative phase is fully restored:
        \begin{align}
            &\text{Case 2: } (-1)^{s_1+s_2+1}\ket{0}_b\ket{g}_1\ket{e}_2\ket{e}_3\ket{\psi}_\mathrm{targ} \rightarrow \ket{0}_b\ket{g}_1\ket{e}_2\ket{e}_3\ket{\psi}_\mathrm{targ},\\
            &\text{Case 4: } (-1)^{s_1+s_2+1}\ket{0}_b\ket{e}_1\ket{g}_2\ket{e}_3\ket{\psi}_\mathrm{targ} \rightarrow \ket{0}_b\ket{e}_1\ket{g}_2\ket{e}_3\ket{\psi}_\mathrm{targ},\\
            &\text{Case 6: } (-1)^{s_1+s_2+1}\ket{0}_b\ket{e}_1\ket{e}_2\ket{e}_3\hat{U}\ket{\psi}_\mathrm{targ} \rightarrow \ket{0}_b\ket{e}_1\ket{e}_2\ket{e}_3\hat{U}\ket{\psi}_\mathrm{targ}.
        \end{align}
    \end{itemize}
\end{itemize}

\section{Derivation of the number of RSB pulses implementing \(M\) \(N\)-controlled gates}
\label{app:deriv_of_successive}

Suppose that we successively implement \(M\) \(N\)-controlled gates with control conditions \(\mathbf{x}_j\) \((1 \le j \le M)\), using the circuits in Fig.~\ref{fig:mc_gate}~(b--e).
We assume that each pair of adjacent multi-controlled gates is implemented such that the second RSB sequence of the first gate and the first RSB sequence of the second gate consist of opposite RSB pulses.
For example, if all gates are implemented using Fig.~\ref{fig:mc_gate}~(d), then the first RSB sequence of each gate consists of RSB \(\pi\) pulses, whereas the second consists of RSB \(-\pi\) pulses.

Consider the implementation of the \(k\)th and \((k+1)\)th gates.
Let \(c_k\) denote the first bit index at which \(\mathbf{x}_k\) and \(\mathbf{x}_{k+1}\) differ, and set \(c_k = N + 2\) if \(\mathbf{x}_k = \mathbf{x}_{k+1}\).

When \(\mathbf{x}_k \neq \mathbf{x}_{k+1}\), the initial \(c_k - 1\) control qubits share identical control conditions.
Therefore, the last \(c_k - 1\) RSB pulses in the second RSB sequence of the \(k\)th gate and the first \(c_k - 1\) RSB pulses in the first RSB sequence of the \((k+1)\)th gate cancel.
Hence, a total of \(2(c_k - 1)\) RSB pulses are eliminated.

When \(\mathbf{x}_k = \mathbf{x}_{k+1}\), all control qubits have identical control conditions, and in this case the rotation gates and the RSB pulses on the last control qubit also cancel.
Therefore, the entire second RSB sequence of the \(k\)th gate and the first RSB sequence of the \((k+1)\)th gate are removed, so that a total of \(2(N+1)=2(c_k-1)\) RSB pulses are eliminated, consistent with the \(\mathbf{x}_k \neq \mathbf{x}_{k+1}\) case.

As a result, the total number of eliminated RSB pulses is
\begin{align}
\label{eq:cancel_sum}
    \sum_{k=1}^{M-1} 2(c_k - 1).
\end{align}
Hence, \(M\) \(N\)-controlled gates can be implemented using
\begin{align}
    2M(N+1) - \sum_{k=1}^{M-1} 2(c_k - 1)
\end{align}
RSB pulses in total.

\section{Detailed numerical results for the 3-CSWAP gate simulation}
\label{app:supp_numerical}

In this section, we present the simulation setup and full numerical results of the 3-controlled SWAP (3-CSWAP) gate discussed in Sec.~\ref{sec:numerical_validation}.

\subsection{Simulation parameters}
\label{app:sim_params}

The numerical simulations model the open-system dynamics of five ion qubits coupled to a single shared motional mode.
The motional Hilbert space is truncated to a Fock cutoff of \(n_{\max} = 10\), and the motional mode is initialized in a thermal state with mean phonon number \(\bar{n} = 0.05\).
We solve the Lindblad master equation using QuTiP's mcsolve routine with 1000 trajectories~\cite{qutip5}.

The Rabi frequency and Lamb--Dicke parameter are set to \(\Omega/2\pi = 0.2\,\mathrm{MHz}\) and \(\eta = 0.1\), respectively, for all RSB and carrier pulses.
Under these conditions, the duration of an RSB \(\pm\pi\) pulse is \(25\,\mu\mathrm{s}\) according to Eqs.~\eqref{eqn:rsb_unitary} and \eqref{eqn:rsb_aux_unitary}, while that of a carrier pulse for \(\hat{R}(\pi/2,\pm\pi/2)\) is \(1.25\,\mu\mathrm{s}\).
We include three error channels---motional heating (\(\hat{L}_h\)), motional dephasing (\(\hat{L}_\phi\)), and Zeeman dephasing (\(\hat{L}_z\))---with rates \(\gamma_h = 1.3 \times 10^{-4}\,\mu\mathrm{s}^{-1}\), \(\gamma_\phi = 5.0 \times 10^{-4}\,\mu\mathrm{s}^{-1}\), and \(\gamma_z = 2.0 \times 10^{-3}\,\mu\mathrm{s}^{-1}\), respectively.
These values are motivated by typical experimental conditions for \({}^{171}\mathrm{Yb}^{+}\) trapped-ion systems~\cite{fang2023realization}.

\subsection{State fidelity analysis}

\begin{figure*}[htb!]
    \centering
    \includegraphics[width=\textwidth]{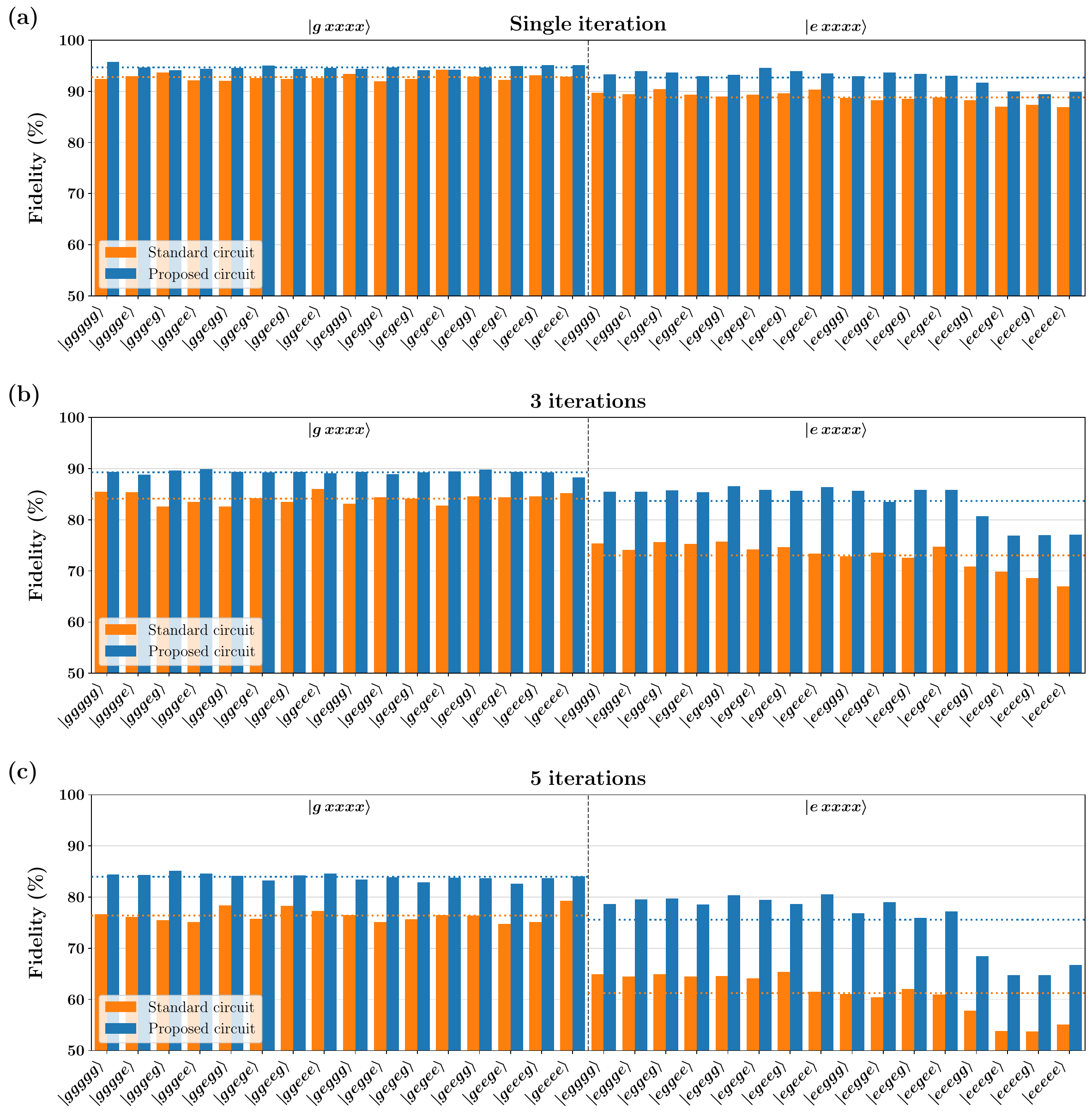}
    \caption{The state fidelities for all 32 computational basis output states of the 3-CSWAP gate, comparing the standard (red) and proposed (blue) circuits.
    (a) Single iteration. (b) 3 iterations. (c) 5 iterations.
    The states are grouped into \(\ket{g\,xxxx}\) and \(\ket{e\,xxxx}\), separated by the dashed vertical line.
    Dotted horizontal lines indicate the group mean fidelity for each scheme.
    The simulation parameters are listed in Appendix~\ref{app:supp_numerical}.}
    \label{fig:fidelity_all_states}
\end{figure*}

Figure~\ref{fig:fidelity_all_states} shows the state fidelities for all 32 computational basis output states, comparing the standard and proposed circuits under the dissipation conditions described above.
The output states are grouped into \(\ket{g\,xxxx}\) (first control qubit in \(\ket{g}\)) and \(\ket{e\,xxxx}\) (first control qubit in \(\ket{e}\)), where each \(x\in\{g,e\}\), separated by the dashed vertical line.
The dotted horizontal lines indicate the group mean fidelity.
Table~\ref{tab:fidelity_summary} summarizes the mean fidelities and standard deviations for each group and the overall average.

\begin{table}[htb!]
\centering
\caption{Summary of state fidelities (\%) for the 3-CSWAP gate and its iterated sequences.
Mean and standard deviation are computed over the 16 states in each group.}
\label{tab:fidelity_summary}
\begin{ruledtabular}
\begin{tabular}{clccc}
\multirow{2}{*}{Iteration} & \multirow{2}{*}{Circuit} & \multicolumn{2}{c}{Group mean $\pm$ std (\%)} & \multirow{2}{*}{Overall (\%)} \\
\cline{3-4}
& & $\ket{g\,xxxx}$ & $\ket{e\,xxxx}$ & \\
\hline
\multirow{2}{*}{1} & Standard & $92.8 \pm 0.6$ & $88.8 \pm 1.1$ & $90.8$ \\
 & Proposed & $94.7 \pm 0.4$ & $92.7 \pm 1.6$ & $93.7$ \\
\hline
\multirow{2}{*}{3} & Standard & $84.2 \pm 1.1$ & $73.0 \pm 2.6$ & $78.6$ \\
 & Proposed & $89.3 \pm 0.4$ & $83.7 \pm 3.6$ & $86.5$ \\
\hline
\multirow{2}{*}{5} & Standard & $76.4 \pm 1.3$ & $61.2 \pm 4.1$ & $68.8$ \\
 & Proposed & $83.9 \pm 0.7$ & $75.6 \pm 5.8$ & $79.7$ \\
\end{tabular}
\end{ruledtabular}
\end{table}

First, the proposed approach consistently outperforms the standard approach across all output states and all iteration counts.
Second, the fidelity gap between the two approaches is particularly pronounced for the \(\ket{e\,xxxx}\) group: for a single iteration, the proposed approach achieves \(92.7\pm1.6\,\%\) compared to \(88.8\pm1.1\,\%\) for the standard approach.
Third, within the \(\ket{e\,xxxx}\) group, the four \(\ket{eee\,xx}\) states (for which all three control qubits are in \(\ket{e}\)) show a slightly larger fidelity drop than the remaining twelve \(\ket{e\,xxxx}\) states.
This difference becomes more pronounced with increasing iteration count, as summarized in Table~\ref{tab:eee_subgroup}.
These behaviors can be attributed primarily to motional heating.
As discussed in Sec.~\ref{sec:numerical_validation}, the longer occupation time in \(\ket{1}_b\) translates into a larger heating-induced error---both for the \(\ket{e\,xxxx}\) group relative to other output states, and for the \(\ket{eee\,xx}\) states relative to other \(\ket{e\,xxxx}\) states. 
An additional contribution arises from motional dephasing.
Because motional dephasing does not commute with Fock-state transitions, increased transition counts for these states lead to enhanced phase-error accumulation.
Nevertheless, the proposed approach consistently undergoes a shorter occupation time in \(\ket{1}_b\) and fewer such transitions than the standard approach for all output states, which accounts for its uniformly higher fidelity.
These two channels reinforce each other, widening the fidelity gap both between the \(\ket{e\,xxxx}\) group and other output states, and between the \(\ket{eee\,xx}\) states and the remaining \(\ket{e\,xxxx}\) states---separations that grow with iteration count as both effects accumulate (Table~\ref{tab:eee_subgroup}).

Therefore, the pulse cancellation mitigates both channels: the shorter occupation time in \(\ket{1}_b\) reduces heating, and fewer Fock-state transitions reduce dephasing.
Consequently, the proposed approach maintains uniformly higher fidelity than the standard approach across all output states.

Finally, the advantage of the proposed approach scales favorably with circuit depth: the overall fidelity improvement grows from \(2.9\,\%\mathrm{p}\) (single iteration) to \(7.9\,\%\mathrm{p}\) (3 iterations) and \(10.9\,\%\mathrm{p}\) (5 iterations).

\begin{table}[htb!]
\centering
\caption{Mean fidelity (\%) comparison between the \(\ket{eee\,xx}\) states (4 states) and the remaining \(\ket{e\,xxxx}\) states (12 states) within the \(\ket{e\,xxxx}\) group.}
\label{tab:eee_subgroup}
\begin{ruledtabular}
\begin{tabular}{clcc}
\multirow{2}{*}{Iteration} & \multirow{2}{*}{Circuit} & \multicolumn{2}{c}{Mean $\pm$ std (\%)} \\
\cline{3-4}
& & $\ket{eee\,xx}$ & Other $\ket{e\,xxxx}$ \\
\hline
\multirow{2}{*}{1} & Standard & $87.4 \pm 0.6$ & $89.3 \pm 0.7$ \\
 & Proposed & $90.2 \pm 1.0$ & $93.5 \pm 0.5$ \\
\hline
\multirow{2}{*}{3} & Standard & $69.1 \pm 1.7$ & $74.3 \pm 1.1$ \\
 & Proposed & $77.9 \pm 1.8$ & $85.6 \pm 0.7$ \\
\hline
\multirow{2}{*}{5} & Standard & $55.1 \pm 1.9$ & $63.2 \pm 1.9$ \\
 & Proposed & $66.1 \pm 1.8$ & $78.7 \pm 1.4$ \\
\end{tabular}
\end{ruledtabular}
\end{table}

\section{Analytic derivation of RSB pulse count for the LCU implementation}
\label{app:deriv_LCU}

In this section, we derive the total number of RSB pulses required to implement the LCU select operator.
We follow the LCU construction introduced in Sec.~\ref{sec: LCU}, where the select operator consists of \(L\) controlled unitaries, using \(N=\lceil \log_{2}{L} \rceil \) ancillary qubits.

Recall that for \(N\) ancillary qubits, the control condition of \(\hat{U}_l\) is \(\mathbf{x}_{l}^{(N)}=x_{N-1}x_{N-2}\dots x_1x_0\), where \(\sum_{j=0}^{N-1} x_j 2^{j} = l\) for \(0 \leq l \leq L-1\).
Let \(c_l^{(N)}\) denote the quantity corresponding to \(c_k\) introduced in Appendix~\ref{app:deriv_of_successive}, with the gate index relabeled from \(k = 1,\dots,M-1\) to \(l = 0,\dots,L-2\) to match the ordering of the select operator in Eq.~\eqref{eqn: SEL}. 
We then define the sequence \(\{c_l^{(N)}\}_{l=0}^{L-2} = (c_0^{(N)}, c_1^{(N)}, \dots, c_{L-2}^{(N)})\).

Since the exact number of RSB pulses for an arbitrary \(L\) is complicated by the presence of a ceiling function in the general expression, we consider the case \(L = 2^N\), which corresponds to the full binary control space, and express the result in terms of \(N\).
With \(L = 2^N\), the index range for \(\{c_l^{(N)}\}\) is fixed as \(l = 0, \dots, 2^N - 2\); we therefore suppress it and write \(\{c_l^{(N)}\}\).
We construct the \((N+1)\) ancillary qubit control strings from the \(N\) qubit strings.
This induces the following recurrence relations:
\begin{align}
    \mathbf{x}_{l}^{(N+1)} &=
    \begin{cases}
    0\,\mathbf{x}_{l}^{(N)}, & 0 \le l \le 2^N-1,\\
    1\,\mathbf{x}_{l-2^N}^{(N)}, & 2^N \le l \le 2^{N+1}-1,
    \end{cases} \\
    c_l^{(N+1)} &=
    \begin{cases}
    c_l^{(N)}+1, & 0 \le l \le 2^N-2,\\
    1, & l=2^N-1,\\
    c_{l-2^N}^{(N)}+1, & 2^N \le l \le 2^{N+1}-2,
    \end{cases}
\end{align}
where the base case is \(\{c_l^{(1)}\}=(c_0^{(1)})=(1)\).
In sequence notation, the recurrence for \( \{ c_l^{(N+1)} \} \) can be expressed compactly as
\begin{align}
\label{eq:recurr_c}
    \{c_l^{(N+1)}\}=\bigl(\{c_l^{(N)}+1\},\,1,\,\{c_l^{(N)}+1\}\bigr).
\end{align}
The total number of eliminated RSB pulses is given by \(S_N := 2\sum_{l=0}^{2^N-2}(c_l^{(N)}-1)\), as defined in Eq.~\eqref{eq:cancel_sum}.
The recurrence relation for \(\{c_l^{(N)}\}\) induces a corresponding recurrence for \(S_N\):
\begin{align}
\label{eq:recurr_S}
S_{N+1}
&= 2\sum_{l=0}^{2^{N+1}-2}\bigl(c_l^{(N+1)}-1\bigr) \nonumber\\
&= 2\left[\sum_{l=0}^{2^N-2}\bigl(c_l^{(N)}+1-1\bigr) + (1-1)
   + \sum_{l=0}^{2^N-2}\bigl(c_l^{(N)}+1-1\bigr)\right] \nonumber\\
&= 4\sum_{l=0}^{2^N-2} c_l^{(N)} \nonumber\\
&= 2S_N +4(2^N-1) \nonumber\\
&= 2S_N + 2^{N+2} - 4.
\end{align}
This yields the closed-form expression:
\begin{align}
    \label{eq:SN_explicit}
    S_N = (N-2)2^{N+1}+4,
\end{align}
valid for all integers \(N \ge 1\).

We prove Eq.~\eqref{eq:SN_explicit} by induction on \(N\).
For the base case \(N=1\), we have \(S_1 = 0\), which agrees with Eq.~\eqref{eq:SN_explicit}.
For the inductive step, assume the formula holds for some \(N \ge 1\).
Substituting into Eq.~\eqref{eq:recurr_S}, we obtain
\begin{align}
S_{N+1} = 2\bigl((N-2)2^{N+1}+4\bigr) + 2^{N+2}-4 = (N-1)2^{N+2} + 4.
\end{align}
This matches the desired form, completing the proof.

We now apply the preceding analysis to determine the total number of RSB pulses for the LCU select operator.
This operator consists of a sequence of \(L=2^N\) distinct \(N\)-controlled gates.
Without pulse cancellation, the baseline number of required pulses is \(L \times 2(N+1) = 2^{N+1}(N+1)\).
From the preceding analysis, a total of \(S_N\) pulses are cancelled.
Therefore, the total number of RSB pulses is given by subtracting \(S_N\) from the baseline:
\begin{align}
2^{N+1}(N+1) - S_N = 2^{N+1}(N+1) - \bigl[(N-2)2^{N+1}+4\bigr] = 3 \cdot 2^{N+1} - 4.
\end{align}

Thus, in terms of the number of controlled gates \(L\), the total number of RSB pulses is \(6L-4\), where \(L = 2^N\).

\bibliography{reference}
\end{document}